\documentclass[a4paper,11pt]{article}	
\pdfoutput=1 
\usepackage{jheppub} 
\usepackage{mathstyle} 
\usepackage{makeidx}
\usepackage{mathrsfs}

\newcommand{\CA}{\mc{A}}

\newcommand{\CD}{\mc{D}}

\newcommand{\CF}{\mc{F}} 
 
\newcommand{\CH}{\mc{H}} 
\newcommand{\CO}{\mc{O}}
\newcommand{\CS}{\mc{S}}
\newcommand{\CZ}{\mc{Z}}

\newcommand{\rc}{{\rm c}}

\newcommand{\dotDelta}{\stackrel{\bullet}{\Delta}}%
\newcommand{\alienpwr}[2]{\dotDelta_{#1}^{\raisebox{-1ex}{\scriptsize #2}}}%
\newcommand{\alien}[1]{\dotDelta_{#1}}%

\renewcommand{\jjj}[1]{}

\makeindex

\title{\boldmath Resurgent structure of 2d Yang-Mills theory on a
  torus}%
\author[a]{Jiashen Chen}%
\author[b]{Jie Gu}%
\author[a,c]{Xin Wang}%
\affiliation[a]{Interdisciplinary Center for Theoretical Study,
University of Science and Technology of China, Hefei, Anhui 230026, China}
\affiliation[b]{School of Physics and Shing-Tung Yau Center\\
  Southeast University, Nanjing 210096, China}%
\affiliation[c]{Peng Huanwu Center for Fundamental Theory, Hefei, Anhui 230026, China}

\emailAdd{ricepud@mail.ustc.edu.cn}
\emailAdd{jie-gu@seu.edu.cn}
\emailAdd{wxin@ustc.edu.cn}

\preprint{USTC-ICTS/PCFT-25-29}

\abstract{We study the resurgent structure of the topological string
  dual to 2d $U(N)$ Yang-Mills on torus.  We find closed form formulas
  for instanton amplitudes up to arbitrarily high instanton orders,
  based on which we propose the non-perturbative partition function
  including contributions from all the real instantons, which is real
  for positive modulus and string coupling.  We also explore complex
  instantons and find two infinite towers of them.  We expect them to
  correspond to BPS states in type II string.}


\begin{document}
\maketitle
\flushbottom

\section{Introduction}


Two dimensional $U(N)$ Yang-Mills theory is an interesting
non-supersymmetric QFT which is exactly solvable yet still non-trivial
\cite{Migdal:1975zg,Rusakov:1990rs,Witten:1991we,Witten:1992xu}.\footnote{See
  \cite{Blau:1993hj,Cordes:1994fc} for instance for good reviews.}
Furthermore, it is known that in the large $N$ limit, it is dual to a
string theory \cite{Gross:1992tu,Gross:1993hu,Gross:1993yt}.  The
partition function of the 2d Yang-Mills factorizes to the product of a
chiral part and an anti-chiral part, and the chiral part can be
interpreted as counting holomorphic maps from the string worldsheet to
the target space of the gauge theory.  However, the true nature of the
string theory is still mysterious when the target space is a generic
Riemann surface of arbitrary genus and when the 't Hooft coupling is 
finite.  See
\cite{Cordes:1994sd,Horava:1995ic,Vafa:2004qa,Aganagic:2004js,Aharony:2023tam,Benizri:2025xmz,Komatsu:2025sqo}
for recent development.

When the target space is a torus, the situation is much more clear.
By relating partition function of the 2d Yang-Mills with that of 4d
BPS black holes and invoking the OSV conjecture \cite{Ooguri:2004zv},
the dual string theory is identified as topological string theory put 
on a double line bundle over an elliptic curve \cite{Vafa:2004qa}. 
For instance, the partition function of the 2d Yang-Mills theory
factorizes in the large $N$ limit, and the chiral component is argued
to be the partition function of the dual topological string theory.
Indeed, it was shown in
\cite{Dijkgraaf:1996iy,Okuyama:2019rqn,Huang:2021zko} that the chiral
partition function of 2d Yang-Mills can be promoted to non-holomorphic
functions which satisfy the BCOV holomorphic anomaly equations
\cite{Bershadsky:1993ta,Bershadsky:1993cx}, a hallmark of the
partition function of topological string theory.

However, it is not known how to formulate the large $N$ duality
precisely when $N$ is finite.  At large but finite $N$, the
factorization is only schematic and it needs to be modified by summing
over RR fluxes over the finite $T^2$
\cite{Vafa:2004qa,Dijkgraaf:2005bp}.  More importantly, while on the
one side the partition function of 2d Yang-Mills is non-perturbatively
well-defined for any finite $N$, on the other side the topological
string partition function is only defined perturbatively \jjj{Added}
in the large $N$ expansion.  One therefore has to find its
non-perturbative completion, taking into account appropriate
non-perturbative corrections of the order $\CO(\re^{-N})$. \jjj{Added}
These are spacetime instantons on the string theory side due to
D-branes, and they correspond to large $N$ instanton corrections on
the 2d YM theory side.  There have been various attempts to find these
non-perturbative corrections
\cite{Lelli:2002gr,Matsuo:2004nn,deMelloKoch:2005rq,Dhar:2006ru,Okuyama:2018clk}.
The most promising proposal is given by Okuyama and Sakai
\cite{Okuyama:2018clk}, based on free fermion formulation of the 2d
Yang-Mills \cite{Minahan:1993np,Douglas:1993wy}.  However, only the
1-instanton amplitude of the proposal has been tested, and its
multi-instanton amplitudes are not only un-tested, \jjj{Modified} but
in fact run into various problems of inconsistency, especially for
non-vanishing $\theta$ angles.


In this paper, we make a new proposal for the full non-perturbative
partition function for the topological string dual to chiral 2d
Yang-Mills, which overcomes the limitations that plague previous
proposals.  Our proposal exploits recent studies of non-perturbative
corrections to topological string partition functions by the powerful
method of resurgence theory \cite{Ecalle}. \footnote{Resurgence theory
  has been used to study 2d Yang-Mills for small $N$
  \cite{Fujimori:2022qij}.  See \cite{Griguolo:2024ecw} for
  perturbative expansion at instanton sectors for 2d Yang-Mills with
  small $N$.}

The resurgence theory is a powerful mathematical framework that
uncovers the intimate relationship between asymptotic perturbative
series and its non-perturbative corrections, which is crucial for
making sense of the asymptotic perturbative series itself.  See
\cite{Sauzin:2007resurgent,Marino:2012zq,Aniceto:2018bis} for useful
reviews.  The resurgence theory claims that in order to incorporate
appropriately non-perturbative corrections, one needs to upgrade the
perturbative power series to trans-series which are subject to the
same set of constraints or equations.
Furthermore, at least a subset of the non-perturbative corrections are
closely related to the perturbative series, and they transform to each
other via Stokes transformations.  These non-perturbative corrections 
are said to be in the resurgent structure of the perturbative series
\cite{Gu:2021ize}.  They are the most important non-perturbative
corrections, and are usually indispensable if one wishes to make sense
of the asymptotic perturbative series. \footnote{There could be
  non-perturbative corrections in different resurgent structures than
  that of the perturbative series, as in the double-well and the
  cosine model of QM, explained for instance in \cite{Dunne:2014uwm}.}

In the case of topological string theory, the resurgence theory has
been applied heavily to uncover non-perturbative corrections to the
partition function.  In particular, it has been proposed
\cite{Couso-Santamaria2016,Couso-Santamaria2015,Couso-Santamaria:2016vwq}
that non-perturbative partition function of topological string, in the
form of a trans-series, also satisfies the BCOV holomorphic anomaly
equations which control the perturbative partition function.  In
addition, when solving the non-perturbative partition function from
the holomorphic anomaly equations, the integration constant, known as
the holomorphic ambiguity, can be fixed completely by exploiting the
Stokes transformation that connect the perturbative partition function
and its non-perturbative corrections.  Here we restrict ourselves to
the most important non-perturbative corrections, those in the
resurgent structure of the perturbative partition function.  Important
progress has been made in this direction for topological string theory
with local Calabi-Yau threefolds \cite{Gu:2022sqc} or simple compact
Calabi-Yau threefolds \cite{Gu:2023mgf} as background
geometry. \footnote{See further development in
  \cite{Marino:2023gxy,Iwaki:2023rst,Marino:2023nem,Alexandrov:2023wdj,Gu:2024wag}
  and \cite{Marino:2024tbx} for a review.} Full non-perturbative
corrections for both single and multi-instanton amplitudes are solved
up to arbitrary instanton orders, and the Stokes transformation of the
perturbative partition function in terms of these instanton amplitudes
are studied.  An important feature of these instanton amplitudes is
that they are functionals of the perturbative partition function whose
moduli are shifted discretely, which is the typical effect of D-brane
insertion \cite{Aganagic:2003qj,Aganagic:2011mi}.  In fact, these
instanton amplitudes are conjectured to be closely related to BPS
states of stable bound states of D-branes in type IIA string on the
same background geometry.  The action of the instanton amplitude is
the central charge of a BPS state
\cite{Drukker:2011zy,Couso-Santamaria2016,Couso-Santamaria2015}, and
the integer Stokes constant that characterizes the associated Stokes
transformation is the BPS multiplicity, first suggested in
\cite{Gu:2021ize,Gu:2022sqc,Gu:2023mgf} with supporting evidence
appearing later in
\cite{Iwaki:2023rst,Gu:2023wum,Douaud:2024khu,Marino:2024yme}.  See
\cite{Bridgeland:2016nqw,Bridgeland:2017vbr,Alim:2021mhp,Bridgeland:2024ecw,Alim:2024dyi,Alim:2015qma,Couso-Santamaria2017a,Li:2025zyr}
also for related development.

In this paper, we apply the same method to find non-perturbative
corrections to the partition function of the topological string dual
to 2d Yang-Mills.
We find closed form formulas for both single and multi-instanton
amplitudes up to arbitrary instanton orders, and they turn out to be
slightly different from the formulas in the usual topological string
\cite{Gu:2022sqc,Gu:2023mgf}.
Our 1-instanton amplitude agrees with that of Okuyama and Sakai
\cite{Okuyama:2018clk}, but our multi-instanton amplitudes are
different.  They pass high precision numerical tests from resurgence
theory, and they overcome the limitations suffered by previous
proposals.  In particular, our formulas are valid for non-vanishing
$\theta$ as well.  Based on these results, we propose the
non-perturbative partition function \jjj{Modification begins}
\eqref{eq:Znps} which take into account contributions from all real
instantons.  Even though it does not seem to be unique on the surface
and depends on an infinite family of real parameters, we further
conjecture that there is a canonical choice of these parameters
i.e.,~\eqref{eq:Zmed}, which should be used to formulate the precise
duality between 2d Yang-Mills on torus and topological string for
large but finite $N$.\jjj{Modification ends.}



\jjj{Modification begins.}  As we mentioned, the non-perturbative
corrections that we have consistently included are large $N$ instanton
corrections for 2d YM theory, or worldsheet instanton corrections for
the dual topological string theory due to D-branes, and these are all
real instantons with real instanton actions.  There could equally be
complex instantons with complex instanton actions.  Even though they
do not play an important role in the non-perturbative partition
function of the chiral 2d YM theory, they do correspond to important
generic D-branes in the dual topological string theory with complex
central charges.  In this paper, we also explore complex instantons,
and we find two infinite towers of complex instanton sectors.  They
should correspond to non-trivial BPS D-brane bound states in type II
string.  We also discuss the wall-crossing behavior of these instanton
sectors. \jjj{Modification ends.}

The structure of the paper is as follows.  In Section~\ref{sc:2dYM} we
review the string interpretation of 2d $U(N)$ Yang-Mills on torus in
the large $N$ limit, the perturbative partition function of the dual
topological string, and the proposal of Okuyama and Sakai
\cite{Okuyama:2018clk} to make the latter non-perturbative.  We
comment on the limitations of their proposal.  In Section~\ref{sc:spe}
we make the duality between the 2d $U(N)$ Yang-Mills on torus and
topological string more concrete by spelling out the special geometry
relations and the singular points in the moduli space, which are
crucial for formulating precisely the holomorphic anomaly equations
that control the dual topological string theory.  Section~\ref{sc:np}
contains the main results of this paper.  We work out the instanton
amplitudes, derive the real non-perturbative partition function, and
finally explore complex instanton sectors.  We conclude in
Section~\ref{sc:con}.

\section{Perturbative and non-perturbative 2d Yang-Mills}
\label{sc:2dYM}

\subsection{String interpretation}

Consider two dimensional Yang-Mills theory with gauge group $G = U(N)$
on a torus $T^2$ with $\theta$ angle turned on.  The action of the
theory is
\begin{equation}
  S = \frac{1}{2g^2_\text{YM}}\int_{T^2} \tr(F\wedge \star F)
    +\theta\tr F.
\end{equation}
Here $\tr$ is the trace in the fundamental representation.  The
partition function can be calculated by summing over representations
$R$ of the gauge group
\cite{Migdal:1975zg,Rusakov:1990rs,Witten:1991we}
\begin{equation}
  Z_N^{\text{YM}} = \sum_R
  \exp\left(-\frac{1}{2}g_{\text{YM}}^2 C_2(R)+\ri\,\theta\,C_1(R)\right),
\end{equation}
where $C_1(R), C_2(R)$ denote the first and the second Casimirs of the
representation.

It was argued in \cite{Gross:1992tu,Gross:1993hu} that in the large
$N$ limit,
\begin{equation}
  N \rightarrow \infty, \quad g_{\text{YM}}\rightarrow 0,\quad
  Ng_{\text{YM}}^2\;\text{fixed},
\end{equation}
the partition function $Z_N^{\text{YM}}$ factorizes to a product of
chiral and anti-chiral components, and the chiral component can be
interpreted as the partition function of a string theory.  By relating
to 4d BPS black hole partition function and applying the OSV
conjecture \cite{Ooguri:2004zv}, it was pointed out in
\cite{Vafa:2004qa} that the string theory is none other than the
topological string theory. \footnote{This is only true when the 2d
  Yang-Mills is put on a torus.  When the target space is a generic
  Riemann surface of arbitrary genus and the 't Hooft coupling is
  finite, the exact string theory that corresponds to the chiral
  partition function is not known.  See for instance
  \cite{Cordes:1994fc,Horava:1995ic,Aganagic:2004js,Komatsu:2025sqo,Benizri:2025xmz}
  for works in this direction.}
More explicitly, the gauge theory partition function factorizes by 
\begin{equation}
  \label{eq:fac}
  Z_N^{\text{YM}} = |\CZ_{\text{top}}(t;g_s)|^2.
\end{equation}
where $\CZ_{\text{top}}(t;g_s)$ is the partition function of the
topological string theory whose target space is the total space of the
double line bundle over the torus $E$
\begin{equation}
  \label{eq:XE}
  X_{E}: \; \CO(1)\oplus \CO(-1) \rightarrow E,
\end{equation}
and which is described by the Gromov-Witten theory of an elliptic
curve \cite{dijkgraaf:1995mirror}.  The string coupling $g_s$ and the
complexified K\"ahler modulus $t$ are given respectively by
\begin{equation}
  g_s = g_{\text{YM}}^2A,\quad t = \frac{1}{2}Ng_s-\ri\theta,
\end{equation}
where $A$ is the torus area, which we will set to one.
The free energy then has the genus expansion
\begin{equation}
  \CF(t;g_s) = \log \CZ_{\text{top}}(t;g_s) = \sum_{g\geq 0}g_s^{2g-2} \CF_g(t).
\end{equation}
The first few terms are\
\begin{equation}
  \CF_0(t) = -\frac{t^3}{6},\quad \CF_1(t) = -\log\eta(q) = \frac{t}{24} -
  \log\prod_{n=1}^\infty (1-q^n),
\end{equation}
with
\begin{equation}
  \label{eq:tau}
  q= \re^{-t} = \re^{2\pi\ri \tau},\quad \tau = \ri t/(2\pi).
\end{equation}
For most part of the paper, we will focus on the region with $\real
t>0$, i.e.~$\imag \tau >0$.

\subsection{Perturbative free energies}
\label{sc:pert}

The free energy has very nice properties. First of all, it was shown
in \cite{dijkgraaf:1995mirror} and proved in \cite{Kaneko1995:msc}
that the free energy $\CF_g(t)$ with $g\geq 2$ is a quasi-modular form
of weight $6g-6$ given by a combination of Eisenstein series.
Second, the free energy $\CF_g(t)$ can be calculated recursively
\cite{Okuyama:2018clk}.  Let us split the holomorphic perturbative
partition function via
\begin{equation}
  \CZ^{\text{top}}(t,g_s) =  \CZ_{01}(t,g_s)\wh{\CZ}^{\text{top}}(t,g_s),
\end{equation}
where
\begin{equation}
  \CZ_{01}(t,g_s) = \exp \left(\CF_0/g_s^2 + \CF_1\right),
\end{equation}
and
\begin{equation}
  \label{eq:FZ}
  \wh{\CZ}^{\text{top}}(t,g_s) = \exp \Bigg(\sum_{g\geq 2}g_s^{2g-2}\CF_g(t)\Bigg) =
  \sum_{n=0}^{\infty} g_s^{2n} \CZ_n^{\text{top}}(t),
\end{equation}
with $\CZ_0^{\text{top}}(t) = 1$.  Then $\CZ_n^{\text{top}}(t)$ satisfies
the recursive relation \cite{Okuyama:2018clk}
\begin{equation}
  \label{eq:Zn-rec}
  \CZ_n^{\text{top}}(t) = h_n^{\text{top}} - \sum_{m=1}^n
  \frac{[D_{-1}+\frac{1}{3}D_3]^{2m}}{(2m)!}
  \CZ_{n-m}^{\text{top}}(t) \cdot (2D_3)^m 1.
\end{equation}
Here $h_n^{\text{top}}$ is defined by the generating series
\begin{equation}\label{eq:hn_recur}
  \sum_{n=0}^\infty g_s^{2n} h_n^{\text{top}}  =
  \exp\left(\sum_{\ell=1}^{\infty} g_s^{2\ell}\frac{e_\ell}{(2\ell)!}\right),
\end{equation}
where
\begin{equation}
  e_\ell = \frac{B_{2\ell+2}}{2\ell+2}2^{-2\ell} D^{2\ell-1}E_{2\ell+2}(q),
\end{equation}
and
\begin{equation}
  D = q\frac{\pd}{\pd q}.
\end{equation}
In addition the derivatives $D_k$ are defined by
\begin{equation}
  D_k = D + \frac{k E_2}{24},
\end{equation}
and we impose in the recursion relation \eqref{eq:Zn-rec} that
$D_{-1}$ only acts on $\CZ^{\text{top}}_{*}$ while $D_3$ only acts on
$1$.  As an example, we find
\begin{equation}
  \label{eq:Z1}
  \CZ^{\text{top}}_{1}(t) = \frac{5E_2^3-3E_2E_4-2E_6}{51840}.
\end{equation}
The free energy $\CF_g(t)$ can then be calculated from the exponential
relation \eqref{eq:FZ}.  

To compute the free energy $\CF_g(t)$ efficiently, we begin with a
high-order, e.g., \jjj{modified} 1000th order, $q$-series expansion of
$e_\ell$. Utilizing the recursive relation
\begin{equation}
  h_n^{\text{top}}
  =\frac{1}{n}\sum_{l=1}^{n}h_{n-l}^{\text{top}} \frac{l\, e_{l}}{(2l)!},
\end{equation}  
derived from \eqref{eq:hn_recur}, the $q$-series expression of
$h_n^{\text{top}} $ can be computed efficiently. Substituting
$h_n^{\text{top}} $ into the recursive relation \eqref{eq:Zn-rec}, we
are able to obtain the \jjj{modified} 1000th-order $q$-series
expression of $\CZ_n^{\text{top}}(t)$ for $n \leq 200$. \footnote{The
  data and the program to generate these data are put on the website
  \texttt{https://github.com/cloud-atlas-47/2dYM-np}.} Since
$\CZ_n^{\text{top}}(t)$ is finitely generated by Eisenstein series, a
sufficiently high-order $q$-series expansion allows one, in principle,
to recover the modular expression of $\CZ_n^{\text{top}}(t)$.
\jjj{Added} For instance, we obtin the full modular expression for
$\CZ_n^{\text{top}}(t)$ up to $n= 60$.
  
Finally, the free energy $\CF_g(t)$ with $g\geq 1$ of topological
string theory has the property that it can be uplifted to a
non-holomorphic version $F_g(t,\bar{t})$, which satisfies the BCOV
holomorphic anomaly equations
\cite{Bershadsky:1993cx,Bershadsky:1993ta}.  In the case of chiral 2d
Yang-Mills, the uplift can be achieved by considering the bosonization
of the free fermion formulation
\cite{Douglas:1993wy,Dijkgraaf:1996iy}.  It was emphasized in
\cite{Okuyama:2019rqn} that in this process, one should distinguish
two kinds of $E_2$, those coming from period integrals over the torus
and those from the so-called propagator, and the uplift is done via
replacing the $E_2$ of the latter type by the almost holomorphic
Eisenstein series
\begin{equation}
  \wh{E}_2(\tau,\bar{\tau}) = E_2(\tau) - \frac{3}{\pi\tau_2},
\end{equation}
where $\tau_2 = \imag\tau$ is the imaginary part of $\tau$, but not
the $E_2$ of the former type.  This is different from the usual case
in topological string theory where all $E_2$ are replaced by
$\wh{E}_2$ when uplifting holomorphic free energy to non-holomorphic
free energy \cite{Huang:2006si,Huang:2013yta}. \footnote{In the chiral
  2d Yang-Mills theory, if all $E_2$ are replaced by $\wh{E}_2$, the
  resulting non-holomorphic free energies do not satisfy holomorphic
  anomaly equations \cite{Okuyama:2019rqn}.}  Once this is taken care
of, it was checked in \cite{Okuyama:2019rqn} and later proved in
\cite{Huang:2021zko} that indeed the non-holomorphic free energies
satisfy the holomorphic anomaly equations
\begin{equation}
  \label{eq:hae}
  \pd_S F_g = \frac{1}{2} \left(\CD^2 F_{g-1} +
    \sum_{h=1}^{g-1}\CD F_h\CD F_{g-h} \right),\quad g\geq 2.
\end{equation}
Here on the right hand side, $\CD$ is the covariant derivative on the
moduli space acting on the free energies via
\begin{equation}
  \CD F_g = D F_g,\;\; \CD^2F_g = (D+S)D F_g.
\end{equation}
On the left hand side, $\pd_S$ is the derivative with respect to the
propagator defined by
\begin{equation}
  \label{eq:S}
  S = \frac{1}{t+\bar{t}} = \frac{1}{4\pi\tau_2} = \frac{E_2 - \wh{E}_2}{12}.
\end{equation}
The derivation is equivalent to the derivative with
respect to $\bar{\tau}$ via
\begin{equation}
  \pd_S = 8\pi\ri (\tau_2)^2\pd_{\bar{\tau}}.
\end{equation}

The genus one free energy is not subject to the holomorphic anomaly
equations \eqref{eq:hae} and its non-holomorphic uplift is
\begin{equation}
  \label{eq:CF1}
  F_1 = \log \frac{\sqrt{2\pi S}}{\eta}.
\end{equation}
In fact, it serves as the initial condition for solving the
holomorphic anomaly equations.  For instance, by direct integration,
we find
\begin{equation}
  F_2 = \frac{5}{24}S^3 - \frac{E_2}{48}S^2 +
  \frac{2E_4-E_2^2}{1152}S + \CF_2(t).
\end{equation}
The holomorphic part $\CF_2(t)$ is the integration constant, also
known as the holomorphic ambiguity, and it can be obtained from
\eqref{eq:FZ} and \eqref{eq:Z1},
\begin{equation}
  \CF_2(t) = \CZ_1^{\text{top}}(t) = \frac{5E_2^3-3E_2E_4-2E_6}{51840}.
\end{equation}
In general, the non-holomorphic free energy $F_g(t,\bar{t})$ is a
polynomial in $S$ of degree $3g-3$
\begin{equation}
  F_g = \sum_{k=0} ^{3g-3} P_n(t) S^n,
\end{equation}
where the holomorphic part $P_0(z)$ is the holomorphic free energy
$\CF_g(t)$.  \jjj{Added} Using this method, we have calculated the
non-holomorphic free energy $F_g$ up to $g=60$.  \footnote{The data
  and the program to generate these data are put on the website
  \texttt{https://github.com/cloud-atlas-47/2dYM-np}.}


\subsection{Non-perturbative proposal}
\label{sc:os}

\jjj{Modification begins.}  It was pointed out in \cite{Vafa:2004qa}
that the factorization \eqref{eq:fac} was only approximately valid for
large but finite $N$, and one had to take into account
non-perturbative corrections $\CO(\re^{-N})$ by summing over RR-fluxes
over the torus, arriving at
\begin{equation}
  \label{eq:fac-sum}
  Z_N^{\text{YM}} =\sum_{n\in\IZ}\CZ_{\text{top}}(t+ng_s;g_s)
  \overline{\CZ}_{\text{top}}(t-ng_s;g_s).
\end{equation}
An even more refined version was proposed in \cite{Dijkgraaf:2005bp}.
It was later argued in \cite{Okuyama:2018clk} that even these
proposals were flawed: e.g., the left hand side of \eqref{eq:fac-sum}
was well-defined for finite $N$ and thus finite $g_s$, while the right
hand side of \eqref{eq:fac-sum} was only defined as asymptotic
perturbative power series of $g_s$. \jjj{Modification ends.}

An attempt to solve this problem was made in \cite{Okuyama:2018clk},
in which process, a non-perturbative version of topological string
free energy was proposed.  The partition function of the 2d Yang-Mills
theory allows a free fermion formulation
\cite{Minahan:1993np,Douglas:1993wy}
\begin{align}
  Z_N^{\text{YM}} =
  &\oint\frac{\rd x}{2\pi\ri x^{N+1}}
    \prod_{n\in\IZ+\frac{N-1}{2}}
    \left(1+x \re^{\ri n\theta}p^{\frac{1}{2}n^2}\right)\nn=
  &\oint\frac{\rd x}{2\pi\ri x^{N+1}}\exp\sum_{k=1}^\infty
    \frac{(-1)^k}{k}x^k
    \begin{cases}
      \vartheta_2(\re^{\ri k \theta},p^k),\quad &\text{even }N,\\
      \vartheta_3(\re^{\ri k \theta},p^k),\quad &\text{odd }N,
    \end{cases} \label{eq:ZNf}
\end{align}
where $p = \re^{-g_s}$.  \jjj{Modification begins} Restricted to the
simple scenario of $\theta = 0$ and assuming that $N$ is even, the
partition function reduces to \jjj{Modification ends}
\begin{equation}
  Z_N^{\text{YM}} = \oint \frac{\rd x}{2\pi\ri x^{N+1}}
  \prod_{n\in\IZ+\frac{1}{2}}\left(1+x p^{\frac{1}{2}n^2}\right).
\end{equation}
It can be written as
\begin{equation}
  \label{eq:ZN-facsum}
  Z_N^{\text{YM}} = \sum_{\substack{N_++N_-=N\\N_\pm\geq 0}} \CZ_{N_+}\overline{\CZ}_{N_-},
\end{equation}
where $\CZ_{N_+}$ and $\CZ_{N_-}$ are defined via the generating
series
\begin{subequations}
  \begin{align}
    &\sum_{N_+=0}^\infty \CZ_{N_+}x^{N_+} = \prod_{n\in\IZ_{\geq 0}+1/2}
      \left(1+x p^{n^2/2}\right),\label{eq:ZN+}\\
    &\sum_{N_-=0}^\infty \overline{\CZ}_{N_-}x^{N_-} = \prod_{n\in-\IZ_{\geq
      0}-1/2}
      \left(1+x p^{n^2/2}\right).\label{eq:ZN-}
  \end{align}
\end{subequations}
Eq.~\eqref{eq:ZN-facsum} looks similar to \eqref{eq:fac-sum}, and it
was proposed in \cite{Okuyama:2018clk} that $\CZ_{N_+}$ be regarded as
the non-perturbative completion of topological string partition
function,
\begin{equation}
  \CZ^{\text{OS}}(t,g_s) = \CZ_{N_+},
\end{equation}
with the dictionary $t = N_+g_s$, in the sense that
\begin{equation}
  \CZ^{\text{OS}}(t;g_s) \sim \CZ^{\text{top}}(t;g_s) \left(1 + \CO(\re^{-1/g_s})\right).
\end{equation}
More explicitly, \jjj{Modified} it was shown in \cite{Okuyama:2018clk}
that
\begin{equation}
  \label{eq:ZOS-phi}
  \CZ^{\text{OS}}(t,g_s)  = \sum_{n=0}^\infty \phi_n(p) \mr{S}^{(+)}\CZ^{\text{top}}(t+ng_s;g_s),
\end{equation}
where $\mr{S}^{(+)}$ is the Borel resummation (see Sec.~\ref{sc:str})
and the functions $\phi_k$ are defined via the generating series
\begin{equation}
  \label{eq:phin}
  \sum_{n=0}^\infty \phi_n(p) y^n =
  \exp\left(\frac{\ri}{2}\sum_{k=0}^\infty \frac{(-y)^k}{k}\vartheta_2(p^k)\right).
\end{equation}
In the leading instanton order
\begin{equation}
  \phi_1(p) = -\frac{\ri}{2}\vartheta_2(p),
\end{equation}
and one has
\begin{equation}
  \CZ^{\text{OS}}(t;g_s) \sim \CZ^{\text{top}}(t;g_s) - \frac{\ri}{2}\vartheta_2(q)
  \CZ^{\text{top}}(t+g_s;g_s) + \ldots
\end{equation}Since in the $g_s\rightarrow 0$ limit
\begin{equation}
  \vartheta_2(p) =
  \sqrt{\frac{2\pi}{g_s}}\vartheta_4(\re^{-4\pi^2/g_s})\approx \sqrt{\frac{2\pi}{g_s}},
\end{equation}
one further has the approximation, 
\begin{equation}
  \label{eq:CZ1}
  \CZ^{\text{OS}}(t;g_s) \sim \CZ^{\text{top}}(t;g_s)  -
  \frac{\ri}{2}\sqrt{\frac{2\pi}{g_s}}\CZ^{\text{top}}(t+g_s;g_s) + \ldots.
\end{equation}
Here
$\CZ^{\text{top}}(t+g_s;g_s)/\CZ^{\text{top}}(t;g_s)\sim
\re^{-t^2/2g_s}$ is of non-perturbative order, representing
corrections from 1-instanton amplitude.  Eq.~\eqref{eq:CZ1} has been
verified in detail by resurgence analysis \cite{Okuyama:2018clk}.

\jjj{Modified} Let us comment that there are several problems with
this proposal.  First of all, even though this proposal includes
corrections from multi-instanton amplitudes to arbitrary high orders,
only the 1-instanton amplitude was tested \cite{Okuyama:2018clk}.
Second, this proposal was derived with the assumption that $N$ is even
and $\theta=0$, which is simply too
restrictive.  
On the one hand, when $N$ is odd, \eqref{eq:ZNf} seems to indicate
that in the generating series \eqref{eq:phin} $\vartheta_2$ should be
replaced by $\vartheta_3(p)$, which nonetheless cannot be
distinguished by the 1-instanton test, as in the $g_s\rightarrow 0$
limit,
\begin{equation}
  \vartheta_2(p) \approx \vartheta_3(p) \approx \sqrt{\frac{2\pi}{g_s}}.
\end{equation}
On the other hand, it was admitted in \cite{Okuyama:2018clk} that
their proposal runs into problems for non-vanishing $\theta$.
Finally, as we will see in Section~\ref{sc:np}, the proposed
non-perturbative partition function has a fictitious non-vanishing
imaginary part when $t$ and $g_s$ are positive.  We will overcome
these problems and make a more consistent proposal for the
non-perturbative partition function.

\section{Special geometry of chiral 2d Yang-Mills}
\label{sc:spe}


In this section, we discuss some important properties of the moduli
space of topological string theory (see \cite{klemm2018b} for instance
for review) and verify that the chiral 2d Yang-Mills theory indeed
enjoys these properties.  We restrict our discussion to the cases
where the moduli space is complex one dimensional, which is
parametrized by a suitable global complex variable $z$. 

The moduli space of topological string theory is a special K\"ahler
manifold.  One of the consequences is that one can define a
non-holomorphic object called the propagator $S(z,\bar{z})$ over the
moduli space from the non-holomorphic genus one free energy by
\begin{equation}
  \label{eq:F1-S}
  \pd_z F_1 = \frac{1}{2}C(z)S + h(z).
\end{equation}
Here $C(z)$ is the Yukawa coupling, which is defined through the
Griffiths transversality of the holomorphic (3,0) form $\Omega$ of the
Calabi-Yau threefold, and which is holomorphic in $z$.  $h(z)$ is also
holomorphic in $z$, but it can be chosen by hand and regarded as a
gauge choice of the theory.

The propagator $S$ has many nice properties.  It satisfies a
differential relation
\begin{equation}
  \label{eq:DS}
  \pd_z S = C(z) \left(S^2 + 2\mf{s}(z)S + \mf{f}(z)\right),
\end{equation}
where both $\mf{s}(z)$, $\mf{f}(z)$ are holomorphic, therefore
generating a differential ring called the BCOV ring
\cite{Hosono:2008np}. It is also related to the Levi-Civita connection
$\Gamma^z_{zz}$ on the moduli space by
\begin{equation}
  \Gamma^z_{zz} = -C(z) (S+\mf{s}(z)).
\end{equation}
Finally, the non-holomorphic free energies $F_g(z,\bar{z})$ of
$g\geq 2$ can be written as polynomials in $S$ whose coefficients are
rational functions in $z$, and they are subject to the BCOV
holomorphic anomaly equations
\cite{Bershadsky:1993cx,Bershadsky:1993ta}
\begin{equation}
  \label{eq:BCOV}
  \pd_S F_g = \frac{1}{2}\left(\CD^2F_{g-1} +\sum_{h=1}^{g-1}\CD_z F_h
    \CD_z F_{g-h}\right),\quad g\geq 2.
\end{equation}
Here $\CD$ is the covariant derivative on the moduli space, acting on
the free energies via
\begin{equation}
  \CD_z F_g = \pd_z F_g,\quad \CD_z^2 F_g = (\pd_z - \Gamma^z_{zz})\CD_z F_g.
\end{equation}

Another important consequence of special K\"ahler manifold is that in
each local open patch of the moduli space one can choose a symplectic
basis of the period integrals $(T,T_D)$ of the holomorphic (3,0) form
$\Omega$, known as the choice of a frame $\Gamma$, so that $T$ is the
flat coordinate on the local patch, and that the basis satisfies the
special geometry relation
\begin{equation}
  T_D = c \frac{\pd \CF_0(T)}{\pd T},
\end{equation}
where $\CF_0$ is the prepotential or the genus zero free energy in the
frame, and $c$ some normalization constant.  Once a frame is chosen,
one can reduce the propagator to its holomorphic limit
\begin{equation}
  \label{eq:S-hol}
  S \rightarrow \CS_\Gamma = -\frac{1}{C(z)}\frac{\pd_z^2 T}{\pd_z T}  -\mf{s}(z),
\end{equation}
and likewise for the non-holomorphic free energies
\begin{equation}
  F_g(z,\bar{z}) \rightarrow \CF^\Gamma_g(z) = F_g(z,S=\CS_\Gamma).
\end{equation}
We use Roman letters to denote non-holomorphic objects and caligraphic
letters to denote the holomorphic limit.

In the case of chiral 2d Yang-Mills theory, by comparing
\eqref{eq:hae} and \eqref{eq:BCOV}, we can choose the global complex
variable
\begin{equation}
  \label{eq:z}
  z = -t,
\end{equation}
and $C(z)=1, \mf{s}(z)=0$ so that $\Gamma^z_{zz} = -S$.  The
propagator $S$ given by \eqref{eq:S} can fit into the definition
\eqref{eq:F1-S} if we make the gauge choice $h(z) = -E_2(q)/24$.
Finally, the propagator $S$ indeed enjoys the differential relation
\begin{equation}
  \pd_zS = -\pd_t S = S^2,
\end{equation}
with simply $\mf{s}(z) = \mf{f}(z) = 0$.

There are two important frames for the chiral 2d Yang-Mills.  The
first frame is the so-called ``\texttt{large radius}'' frame, where we
choose $t$ as the flat coordinate, and the dual period integral is
\begin{equation}
  t_D = t^2/2,
\end{equation}
so that they satisfy the special geometry relation
\begin{equation}
  t_D = -\frac{\pd \CF_0}{\pd t}.
\end{equation}
The period modulus $\tau$ defined by \eqref{eq:tau} can be written as
\begin{equation}
  \tau = \frac{\ri}{2\pi}\frac{\pd t_D}{\pd t} =
  -\frac{\ri}{2\pi}\frac{\pd^2 \CF_0}{\pd t^2}.
\end{equation}
Note that here the relationship between the flat coordinate $t$ and
the global variable $z$ known as the mirror map is trivial, given by
\eqref{eq:z}, unlike the usual case in topological string.  The
holomorphic free energies calculated in Section~\ref{sc:pert} are in
fact the holomorphic limit of the non-holomorphic free energies in the
\texttt{large radius} frame.  Indeed, we find by \eqref{eq:S-hol} that
in this limit
\begin{equation}
  \label{eq:SLR}
  S\rightarrow \CS_{\text{LR}} = \frac{\pd_t^2 t}{\pd_t t}  = 0,
\end{equation}
which can be equally obtained by sending $\bar{t}\rightarrow \infty$,
and the non-holomorphic free energies $F_g(t,\bar{t})$ reduce to
$\CF_g(t)$.

The other important frame is the so-called ``\texttt{conifold}''
frame, where we exchange the roles of the two period integrals.  We
choose
\begin{equation}
  t_c = \frac{t^2}{2}
\end{equation}
as the flat coordinate and $t_{c,D} = t$ as the dual period integral.
The \texttt{conifold} frame is related to the \texttt{large radius}
frame by an $S$-transformation: this is seen clearly from $\tau$
\begin{equation}
  \label{eq:S-trfs}
  \tau_{\rc} := -2\pi\ri\frac{\pd t_{c D}}{\pd t_c} = -\frac{1}{\tau}.
\end{equation}
The holomorphic limit of the propagator is
\begin{equation}
  \label{eq:Sc}
  S\rightarrow \CS_{\rc} = \frac{\pd_t^2 t_c}{\pd_t t_c}  = \frac{1}{t},
\end{equation}
which corresponds to sending $\bar{t} \rightarrow 0$.  Equivalently,
this limit can be obtained by performing the $S$-transformation
\eqref{eq:S-trfs} on the Eisenstein series $E_2(\tau), \wh{E}_2(\tau)$
in \eqref{eq:S}, taking in account that
\begin{equation}
  E_2(\tau) = \tau^{-2}E_2(-1/\tau) + \frac{12}{t},\quad \wh{E}_2(\tau) =
  \tau^{-2}\wh{E}_2(-1/\tau),
\end{equation}
and then replacing $\wh{E}_2$ back by $E_2$.  Similarly, the
holomorphic limit of the free energy in the \texttt{conifold} frame is
obtained by performing the $S$-transform on all the Eisenstein series
followed by the replacement $\wh{E}_2\rightarrow E_2$ or by simply
sending $\bar{t}\rightarrow 0$.

There are also two singular points in the moduli space.  The first
singular point is the ``large radius'' point located at
\begin{equation}
  t \rightarrow  \infty,
\end{equation}
or $q= \re^{-t}\rightarrow 0$, corresponding to the large $t$ regime of
\cite{Okuyama:2019rqn}.  Near the large radius point, the holomorphic
limit of the free energy $\CF_g$ in the \texttt{large radius} frame
has the asymptotic behavior
\begin{equation}
  \CF^{\text{LR}}_g  = \CF_g\sim \frac{q^2}{(g-1)(2g-3)!} + \CO(q^3),\quad g\geq 2.
\end{equation}

The second singular point is the ``conifold'' point located at
\begin{equation}
  t = 0,
\end{equation}
and it corresponds to the small $t$ regime of \cite{Okuyama:2019rqn}.
Near the conifold point, the holomorphic limit of the free energy
$\CF_g^{\rc}$ in the conifold frame has the asymptotic
behavior\footnote{If we choose the naive non-holomorphic uplift of
  $\CF_g$, simply replacing all $E_2$ by $\wh{E}_2$, the
  \texttt{conifold} frame free energy $\CF^{\rc}_g$ would then not
  display gapped asymptotic behavior near $t=0$.  This is another
  piece of evidence that the naive non-holomorphic uplift of free
  energy does not work.}  \cite{Okuyama:2019rqn}
\begin{equation}
  \label{eq:gap}
  \CF^{\rc}_g \sim \frac{(-1)^g16\sqrt{2}
    \pi^{2g}(4g-5)!B_{2g}(1/2)}{2^{2g}(2g)!(2g-3)!t_c^{2g-3/2}}
  +\CO(\re^{-4\pi^2/t}),\quad g\geq 2,
\end{equation}
This is similar to the gap condition of the conifold frame free energy
of topological string theory near a conifold singular point
\cite{Ghoshal:1995wm}, although the leading exponent of the flat
coordinate and the leading coefficient are different.




\section{Instanton amplitudes and resurgent structure}
\label{sc:np}

\subsection{General strategy}
\label{sc:str}

We study in the following non-perturbative corrections to the free
energy of the topological string theory dual to chiral 2d Yang-Mills.
We first describe our general strategy.

There could be various instanton sectors with different instanton
actions $\CA_\alpha$ whose contributions to the non-perturbative
corrections, also called the instanton amplitudes, are of the order
$\re^{-\CA_\alpha/g_s}$.  Here we demand that the different instanton
sectors are such that their instanton actions as complex numbers are
not colinear, i.e.
\begin{equation}
  \arg  \CA_\alpha - \arg \CA_\beta \not\in 2\pi\IZ,\quad
  \forall\alpha\neq \beta.
\end{equation}
In addition, in an instanton sector with action $\CA$, there could be
multi-instanton amplitudes $F^{(n)}$ of the order $\re^{-n\CA/g_s}$
with $n\geq 1$ so that the complete non-perturbative correction from
this instanton sector takes the form of a trans-series,
\begin{equation}
  \label{eq:F-np}
  F^{\text{np}} = F^{(0)} + \sum_{n=1}^\infty F^{(n)}.
\end{equation}
Here $F^{(0)}$ is the power series of the perturbative free energy
\begin{equation}
  F^{(0)} = \sum_{g=0}^{\infty} g_s^{2g-2} F_g,
\end{equation}
and the instanton amplitude $F^{(n)}$ at order $n$ takes the form
\begin{equation}
  F^{(n)} = \re^{-n\CA/g_s}g_s^{\nu_n}\sum_{k=0}^\infty g_s^k F_k^{(n)}.
\end{equation}

Our general strategy to find instanton amplitudes is based on two
crucial assumptions following
\cite{Couso-Santamaria2016,Couso-Santamaria2015,Gu:2022sqc,Gu:2023mgf}.
The first assumption is that the trans-series \eqref{eq:F-np}
including all instanton contributions from a particular instanton
sector also satisfies the BCOV holomorphic anomaly equations.  We can
then expand the holomorphic anomaly equations in terms of two small
parameters $g_s$ and $\re^{-\CA/g_s}$ leading to recursion relations
for $F_k^{(n)}$, which can be solved by direct integration
\cite{Couso-Santamaria2016,Couso-Santamaria2015}.  In this process,
just like solving the perturbative free energy $F_g$, there will also
be holomorphic ambiguities in the instanton amplitudes.
To fix them, we make the second assumption that the instanton
amplitudes are closely related to the perturbative free energy in
accord with the expectation of the resurgence theory \cite{Ecalle}.
What this second assumption means is the following.


The trans-series \eqref{eq:F-np} is a formal object as both the
perturbative power series $F^{(0)}$ and the power series in each
instanton amplitude $F^{(n)}$ are divergent power series of the
1-Gevrey type, whose coefficients grow factorially fast,
\begin{equation}
  F_g\sim (2g-2)!,\quad F_k^{(n)}\sim k!.
\end{equation}
In general given a divergent series of 1-Gevrey type
\begin{equation}
  \varphi(z) = \sum_{n=0}^\infty a_n z^n,\quad a_n\sim n!,
\end{equation}
we can resum it and convert it to a finite number by the means of
Borel resummation.  The Borel resummation is the Laplace transform of
the Borel transform
\begin{equation}
  \label{eq:Borel-sum}
  \mr{S}\varphi(z) = \frac{1}{z}\int_0^{\re^{\ri\arg z}\infty}
  \mc{B}[\varphi](\zeta)\re^{-\zeta/z}\rd\zeta,
\end{equation}
which in turn is a convergent power series defined by
\begin{equation}
  \mc{B}[\varphi](\zeta) = \sum_{n=0}^\infty\frac{a_n}{n!}\zeta^n.
\end{equation}
The Borel resummation is well-defined if the Borel transform as the
sum of a convergent series can be analytically continued along the
direction of $\arg z$ to infinity.  This is possible if along this
direction there are no singular points of the Borel transform, also
known as Borel singularities.  For each Borel singularity $\CA_\alpha$
with $\vartheta_\alpha = \arg\CA_\alpha$, define a Stokes ray
$\rho_\alpha = \re^{\ri\vartheta_\alpha}\IR_+$ in the complex $z$
plane.  Then the Borel resummation defines a continuous function in
disjoint cones in the complex $z$ plane, which are separated by Stokes
rays.  The discontinuity of the function across a Stokes ray
$\rho_\alpha$ is given by the discrepancy of the pair of lateral Borel
resummations, whose integration contours sandwich the Stokes ray, i.e.
\begin{equation}
  \text{disc}_{\vartheta_\alpha}\varphi(z) = \mr{S}^{(+)} \varphi(z) -
  \mr{S}^{(-)} \varphi(z),
\end{equation}
with
\begin{equation}
  \label{eq:latsum}
  \mr{S}^{(\pm)} \varphi(z) =
  \frac{1}{z}\int_0^{\re^{\ri\vartheta_\alpha\pm\ri 0^\pm}\infty}
  \mc{B}[\varphi](\zeta)\re^{-\zeta/z}\rd\zeta.
\end{equation}
This is known as the Stokes discontinuity, and it is of the order of
magnitude $\re^{-\CA_\alpha/z}$, indicative of possible
non-perturbative corrections.  In fact, by the resurgence theory, the
Borel singularity $\CA_\alpha$ that gives rise to the Stokes ray
$\rho_\alpha$ is the action of an instanton sector.  In the simple
case where there are no multi-instantons (only one Borel singularity
on the Stokes ray), the Stokes discontinuity is simply given by the
Borel resummation of the associated instanton amplitude
\begin{equation}
  \label{eq:disc}
  \text{disc}_{\vartheta_\alpha}\varphi(z) = \CS_{\alpha} \mr{S}^{(-)}\varphi^{(\alpha)}(z),
\end{equation}
with
\begin{equation}
  \varphi^{(\alpha)}(z) = \re^{-\CA_\alpha/z}z^{-\nu_\alpha}\sum_{k=0}^\infty a_k^{(\alpha)}z^k,
\end{equation}
up to a proportionality constant $\CS_{\alpha}$ known as the Stokes
constant.  The formula of Stokes discontinuity \eqref{eq:disc} also
implies the famous resurgence relation: the coefficients of the
instanton amplitudes control the asymptotic behavior of the
perturbative coefficients
\begin{equation}
  \label{eq:res-rel}
  a_n \sim \frac{\CS_\alpha}{2\pi\ri}\frac{\Gamma(n+\nu_\alpha)}{\CA^{n+\nu_\alpha}}
  \sum_{k=0}^\infty \frac{a_k^{(\alpha)}\CA_\alpha^k}{\prod_{j=1}^k(n+\nu_\alpha-j)}.
\end{equation}
The formula of Stokes discontinuity \eqref{eq:disc} and resurgence
relation \eqref{eq:res-rel} allow us to extract information of the
instanton amplitudes from the perturbative power series alone.

If there are different instanton sectors, their contributions add up
in the resurgence relation.  Obviously, the instanton sector whose
instanton action has the smallest absolute value plays the dominant
role.  On the other hand, in the case of resonant resurgence where
instanton sectors arise in pairs with opposite instanton actions,
their contributions to odd power perturbative coefficients cancel, and
one gets
\begin{equation}
  \label{eq:z2n}
  \varphi(z) =\sum_{n=0}^{\infty} a'_n z^{2n},\quad a'_n \sim (2n)!,
\end{equation}
whose remaining coefficients satisfy the resurgence relation
\begin{equation}
  \label{eq:res-rel2}
  a'_n \sim \frac{\CS_\alpha}{\pi\ri}\frac{\Gamma(2n+\nu_\alpha)}{\CA_\alpha^{2n+\nu_\alpha}}
  \sum_{k=0}^\infty \frac{a_k^{(\alpha)}\CA_\alpha^k}{\prod_{j=1}^k(2n+\nu_\alpha-j)}.
\end{equation}
This is what we will encounter in topological string.

For generic cases where there are multi-instanton contributions with
actions $\ell \CA$ $(\ell=2,3,\ldots)$, the impact of the higher order
instanton amplitudes is more complicated, although it can still be
described in a precise way as discussed in more detail in
Section~\ref{sc:m-inst}.  In the special case where the
multi-instanton amplitudes truncate to finite powers, their
contributions to the formula of Stokes discontinuity also simply add up, just like
contributions from different instanton sectors.

The instanton sectors which contribute to the Stokes discontinuity of
the perturbative power series are said to form the resurgent structure
from the perturbative series \cite{Gu:2021ize}.  They may constitute
only a subset of all possible non-perturbative contributions, but they
are the most important ones.  Here, we will only be interested in this
type of non-perturbative contributions to the topological string
theory dual to chiral 2d Yang-Mills.

These non-perturbative contributions to topological string theory have
been studied in other examples, for instance when the background
geometry is a local Calabi-Yau threefold \cite{Gu:2022sqc} or a simple
compact Calabi-Yau threefold \cite{Gu:2023mgf} (see
\cite{Marino:2023nem,Alexandrov:2023wdj,Gu:2024wag} also for
generalization to real topological string and refined topological
string as well as to other observables in topological string.).
Several observations were made which were then conjectured to be true
for generic Calabi-Yau background.  We will test these conjectures for
the background geometry $X_E$, the double line bundle over an elliptic
curve \eqref{eq:XE}.  

It was demonstrated in \cite{Gu:2022sqc,Gu:2023mgf} that the resurgent
assumption puts very strong constraints on possible instanton sectors.
The instanton actions are in fact all integral periods of the
Calabi-Yau background
\cite{Drukker:2011zy,Couso-Santamaria2016,Couso-Santamaria2015}, each
of which can be used to define the flat coordinate of a symplectic
frame, known as the $A$-frame of the instanton sector.  In addition,
given an instanton sector with action $\CA$, the holomorphic limit of
the instanton amplitudes in the $A$-frame simplify greatly, reducing
to truncated power series of finite degrees, and they can be entirely
determined by the resurgent properties of the perturbative free
energy.  In turn, these $A$-frame instanton amplitudes can serve as
boundary conditions to fix completely the holomorphic ambiguity of the
instanton amplitudes as solutions to the holomorphic anomaly
equations. 

\begin{figure}
  \centering
  \subfloat[$t=4\pi$]{\includegraphics[height=4cm]{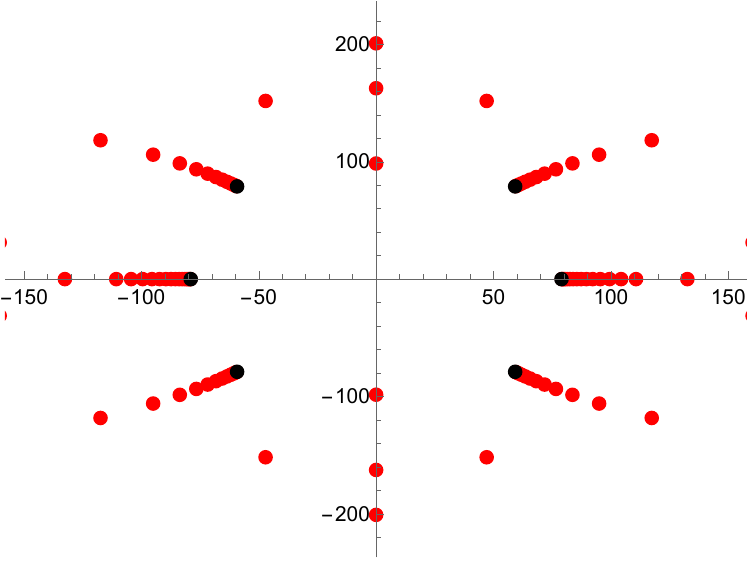}}\hspace{4ex}
  \subfloat[$t=\pi$]{\includegraphics[height=4cm]{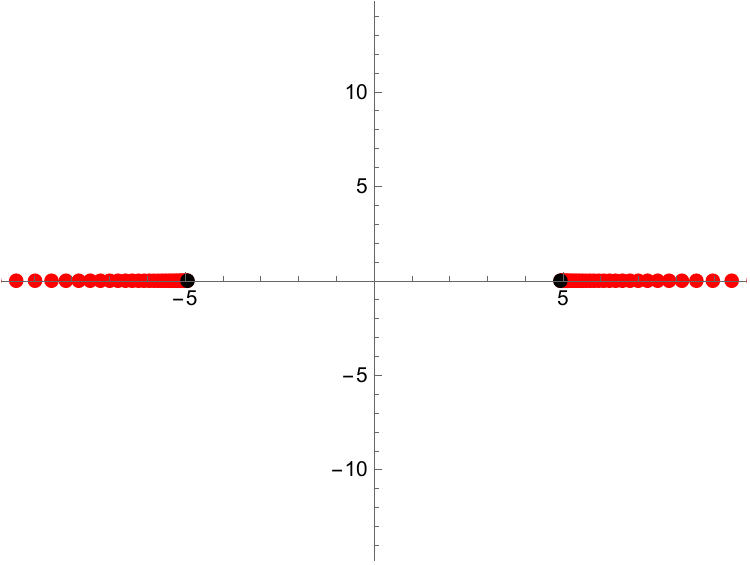}}
  \caption{Borel singularities of perturbative free energy in the
    \texttt{large radius} frame.  We use perturbative series truncated
    to 200 terms, and use Pad\'e approximant to mimic the analytic
    continuation of the Borel transform.  The singular points (red) of
    the approximation would condense to branch cuts if the truncation
    is pushed to infinity.  At $t=4\pi$ (a), the branch points (black)
    have charges $\pm (1,0,0), \pm (1,2,2), \pm (1,-2,2)$. At $t=\pi$
    (b), the branch points (black) have charges $\pm (1,0,0)$.}
  \label{fig:brl}
\end{figure}

We show in this section that all these observations still hold true
for the topological string theory dual to the chiral 2d Yang-Mills,
although there are subtle differences.  First of all, simple plots of
Borel singularities of perturbative free energy as in
Figs.~\ref{fig:brl} show that the instanton actions are periods of the
form
\begin{equation}
  \label{eq:A-gamma}
  \CA_\gamma = \alpha t^2/2 + \beta 2\pi\ri\, t +\delta \pi^2,
\end{equation}
with the charge vector $\gamma = (\alpha,\beta,\delta)\in\IZ^3$.  The
dominant instanton sectors have actions
\begin{equation}
  \label{eq:Ac}
  \pm\CA_c = \pm t^2/2,
\end{equation}
and we call them real instantons as the actions are real for $t>0$.
In Section~\ref{sc:bdy} we solve completely both the 1-instanton and
multi-instanton amplitudes in the real instanton sectors in the
$A$-frame by exploiting the resurgence relations and the formula of
Stokes discontinuity.  These formulas are valid even when $\theta$ is
turned on so that $t$ is complex.
Using these as boundary conditions, we fix the holomorphic ambiguities
of the trans-series solutions to the holomorphic anomaly equations,
first for 1-instanton amplitude in Section~\ref{sc:1-inst}, and then
for multi-instanton amplitudes in Section~\ref{sc:m-inst}.  We check
these solutions by taking them to non-$A$-frames and comparing with
numerical calculation of the Stokes discontinuity of the perturbative
free energy.

Based on these results, we write down in Section~\ref{sc:med} the
non-perturbative topological string partition function which takes
into account non-perturbative corrections from all the real instantons
and which is real for positive $t$ and $g_s$.  There is in fact an
infinite family of them, and the simplest member of it is the medium
Borel resummation of the perturbative partition function.  We also
compare with the proposal of Okuyama and Sakai \cite{Okuyama:2018clk}
summarized in Section~\ref{sc:os}.

Finally, in Section~\ref{sc:BPS} we study the spectrum of complex
instanton sectors with actions different from \eqref{eq:Ac}.  We find
two infinite towers of complex instantons as well as a pair of
additional complex instantons for which the 1-instanton amplitude
solved in Section~\ref{sc:1-inst} works universally.  We expect they
correspond to BPS bound states of D-branes in type II string.  Their
wall-crossing behavior is also discussed.


\subsection{Boundary conditions}
\label{sc:bdy}

We first explore instanton amplitudes for the real instantons in the
$A$-frame, which is nothing else but the \texttt{conifold} frame.
Recall that the the free energy $\CF_g^{\rc}$ in the conifold frame
displays the gap-like asymptotic behavior \eqref{eq:gap}.
In the leading order, it reads
\begin{equation}
  \label{eq:gapL}
  \CF_g^{\rc} \sim
  \frac{\Gamma(2g-\frac{3}{2})}{\CA_c^{2g-\frac{3}{2}}}
  \sqrt{\frac{2}{\pi}}\left(1+\CO(1/2)^{2g}\right),\quad
  g\geq 2,
\end{equation}
with
\begin{equation}
  \CA_c = t_c = t^2/2.
\end{equation}
By comparing wth the resurgence relation \eqref{eq:res-rel2}, one
concludes that the dominant instanton sectors are indeed the pair with
instanton actions $\pm \CA_c$ \cite{Okuyama:2018clk}.  Furthermore, in
the $A$-frame, denoted by the subscript $\CA$, the 1-instanton
amplitude is simply\footnote{The prefactor $g_s^{-2}$ is to match the
  leading exponent of the perturbative free energy.},
\begin{equation}
  \label{eq:CF1A}
  \CF^{(1),\rc}_\CA = g_s^{-2} \pi\ri\sqrt{\frac{2}{\pi}}g_s^{3/2}\re^{-\CA_c/g_s},
\end{equation}
which has only one term in the power series.  Note that it is
different from the usual case in topological string where 1-instanton
amplitudes truncate to two terms in the $A$-frame
\cite{Pasquetti2010}.

On the other hand, the gap-like asymptotic behavior \eqref{eq:gap} has
subleading contributions other than \eqref{eq:gapL}, which implies
possible multi-instanton amplitudes.  To uncover these higher order
instanton amlitudes, we calculate the Borel resummation of the
following divergent power series
\begin{equation}
  \CF^{\text{gap}}(g_s) =
  \sum_{g=2}^\infty a_g g_s^{2g-2},
\end{equation}
that describes the gap condition, where $a_g$ are the coefficients of
the gap condition \eqref{eq:gap}, given by
\begin{equation}
  a_g = \frac{(-1)^g16\sqrt{2}\pi^{2g}(4g-5)! B_{2g}(1/2)}{2^{2g}(2g)!(2g-3)!}.
\end{equation}

Let $z = g_s/t_c=g_s/\CA_c$, the divergent power series reads
\begin{align}
  \CF^{\text{gap}}(g_s) =
  &\frac{g_s^{1/2}}{t_c}\sum_{g=2}^\infty a_g
    z^{2g-\frac{5}{2}}\nn  =
  &\frac{g_s^{1/2}}{t_c}\sum_{g=2}^\infty
    \frac{a_g}{(2g-\frac{5}{2})!}
    z^{2g-\frac{5}{2}}\int_0^\infty\re^{-\zeta}\zeta^{2g-\frac{5}{2}}\rd
    \zeta\nn\sim
  &\frac{g_s^{1/2}}{t_c}\frac{1}{z}\int_0^{\re^{\ri\vartheta}\infty}
    \re^{-\zeta/z}\sum_{g=2}^\infty\frac{a_g}{(2g-\frac{5}{2})!}\zeta^{2g-\frac{5}{2}}\rd\zeta,\quad
    \vartheta =\arg z.
\end{align}
Here in the third step we illegally exchange the orders of integration
and summation, which is the reason that we obtain in the end a
convergent integral, which is nothing else but the Borel resummation
of the original divergent power series.  In other words, we find
\begin{equation}
  \label{eq:gap-Borel}
  \mr{S} \CF^{\text{gap}}(g_s) = g_s^{-1/2}\int_0^{\re^{\ri\vartheta}\infty}
  \re^{-\zeta \CA_c/g_s}\sum_{g=2}^\infty\frac{a_g}{(2g-\frac{5}{2})!}\zeta^{2g-\frac{5}{2}}\rd\zeta.
\end{equation}

To evaluate the Borel resummation, we use the fact that
\begin{equation}
  \label{eq:B2g}
  B_{2g}(1/2) 
  = (-1)^g \frac{2(2g)!}{(4\pi)^{2g}}
  \sum_{\ell=0}^\infty\left(\frac{1}{(\ell+1/2)^{2g}}-\frac{1}{(\ell+1)^{2g}}\right).
\end{equation}
This allows us to derive the following key identity
\begin{equation}
  \label{eq:key}
  \sum_{g=2}^\infty\frac{a_g}{(2g-\frac{5}{2})!}\zeta^{2g-\frac{5}{2}} =
  -\sqrt{\frac{2}{\pi}}\left(\frac{\pi^2}{12\zeta^{1/2}} +
    \frac{1}{2\zeta^{5/2}} - \frac{\pi}{2\zeta^{3/2}\sin(\pi\zeta)}\right),
\end{equation}
which gives the analytic continuation of the Borel transform inside
the formula of Borel resummation \eqref{eq:gap-Borel}.  The latter can
then be written as
\begin{equation}
  \mr{S} \CF^{\text{gap}}(g_s) = -\sqrt{\frac{2}{\pi g_s}}
  \int_0^{\re^{\ri\vartheta}\infty}\re^{-\zeta \CA_c/g_s}
  \left(\frac{\pi^2}{12\zeta^{1/2}}+\frac{1}{2\zeta^{5/2}}
    -\frac{\pi}{2\zeta^{3/2}\sin(\pi\zeta)}\right)
  \rd\zeta.
\end{equation}
As usually happens in Borel resummation, the integral above is
obstructed when the integrand has singular points along the
integration contour.
This happens when
\begin{equation}
  \vartheta = 0   \;\Leftrightarrow\;   \arg g_s = \arg \CA_c,
\end{equation}
and the integration is along the positive real axis.  The Stokes
discontinuity is defined by
\begin{equation}
  \text{disc}_0 \CF^{\text{gap}}(g_s) =
  -\sqrt{\frac{2}{\pi g_s}}\int_{\mc{H}}
  \re^{-\zeta \CA_c/g_s}
  \left(\frac{\pi^2}{12\zeta^{1/2}}+\frac{1}{2\zeta^{5/2}}-\frac{\pi}{2\zeta^{3/2}\sin(\pi\zeta)}\right)
  \rd\zeta,
\end{equation}
where $\mc{H}$ is the Hankel type integration contour that comes from
$+\infty$ below the positive real axis, half-circles the origin, and
goes away to $+\infty$ above the positive real axis.  The integrand
has a string of poles at\footnote{Note that the integrand is
  \emph{not} singular at the origin, as it is the analytic
  continuation of the Borel transform of $\CF^{\text{gap}}(g_s)$,
  which is a convergent series with a positive radius of convergence.}
\begin{equation}
  \zeta = n, \quad n = 1,2,3,\ldots
\end{equation}
on the positive real axis.  The Stokes discontinuity is then evaluated
by summing up residues of all these poles, and we arrive at
\begin{equation}
  \text{disc}_0 \CF^{\text{gap}}(g_s) =
  \ri\sqrt{\frac{2\pi}{g_s}}\sum_{n=1}^\infty\frac{(-1)^{n-1}}{n^{3/2}}\re^{-n\CA_c/g_s}.
\end{equation}
This indicates that indeed there are multi-instanton contributions
whose actions are
\begin{equation}
  n\CA_c =  nt_c = \frac{nt^2}{2},
\end{equation}
and whose instanton amplitudes in the $A$-frame are
\begin{equation}
  \label{eq:CFnA}
  \CF^{(n)}_{\CA_c} = \mu_n\re^{-n\CA_c/g_s},\quad \mu_n=\ri\sqrt{\frac{2\pi}{g_s}}\frac{(-1)^{n-1}}{n^{3/2}},
\end{equation}
all of which truncate to a single term, in accord with our
expectation.  Note again that they are slightly different from the
usual case in topological string, where in the $A$-frame all
multi-instanton amplitudes truncate to two terms \cite{Gu:2022sqc}.

\subsection{1-instanton amplitudes}
\label{sc:1-inst}

To find the instanton amplitudes for the $\pm\CA_c$ sectors in a
non-$A$-frame, we solve the trans-series solutions to the holomorphic
anomaly equations or HAEs \eqref{eq:hae}.  We solve 1-instanton
amplitude in this subsection and multi-instanton amplitudes in the
next.  In this and the next subsections, we drop the subscript $c$ in
the action $\CA$, since the solutions we find are universal and apply
for other instanton sectors as well, as we will see in
Section~\ref{sc:BPS}.

Let us define
\begin{equation}
  \wt{F} = \sum_{g\geq 1}g_s^{2g}F_g,\quad \wh{F} = \sum_{g\geq 2}g_s^{2g}F_g.
\end{equation}
The HAEs \eqref{eq:hae} can be written in the master form
\begin{equation}
  \label{eq:hae-master}
  \pd_S \wh{F} = \frac{1}{2}g_s^2\CD^2 \wt{F}+ \frac{1}{2} (\CD\wt{F})^2.
\end{equation}
By our first assumption, the trans-series \eqref{eq:F-np} should also
satisfy the HAE \eqref{eq:hae-master} with
\begin{equation}
  \wh{F}^{(\ell)} = \wt{F}^{(\ell)} = g_s^2 F^{(\ell)},\quad \ell\geq 1.
\end{equation}
Expanding up to the 1-instanton order, we find the following linear
equation for $F^{(1)}$,
\begin{equation}
  \label{eq:hae-F1}
  \pd_S F^{(1)} = \frac{g_s^2}{2}\CD^2 F^{(1)} + \CD \wt{F}^{(0)} \CD
  F^{(1)}.
\end{equation}
It was found in \cite{Gu:2022sqc} that
\begin{equation}
  \label{eq:F1-sol}
  \text{const.}\exp\left(-\Phi^{(n)}/g_s\right)
\end{equation}
satisfies the linear equation \eqref{eq:hae-F1}.  Here the function
$\Phi^{(n)}$ in the exponent is,
\begin{equation}
  \label{eq:Phi-n}
  \Phi^{(n)} = \frac{1}{g_s\mb{D}}\left(1-\re^{-n g_s\mb{D}}\right)G
  = \sum_{k=1}^\infty \frac{(-g_s)^{k-1}n^k}{k!}\mb{D}^{k-1}G
\end{equation}
with the derivative $\mb{D}$,
\begin{equation}
  \mb{D} = \CD\CA (S - \CS_\CA) \CD,
\end{equation}
and $\CS_\CA$ the holomorphic propagotor in the $A$-frame.  The
function $G$ is
\begin{equation}
  G = \CA + \mb{D}\wt{F}^{(0)}.
\end{equation}
To fix the normalization constant and the exponent $n$, we evaluate
the instanton amplitude in the holomorphic limit of the $A$-frame.
This is realized by
\begin{equation}
  S\rightarrow \CS_\CA,
\end{equation}
where the derivative $\mb{D}$ simply vanishes, and the expoential in
\eqref{eq:F1-sol} reduces to $\re^{-n\CA/g_s}$.
By comparing with \eqref{eq:CF1A}, we can fix the normalization
constant and the exponent, and the 1-instanton amplitude is
\begin{equation}
  \label{eq:F1}
  F^{(1)} = \ri\sqrt{\frac{2\pi}{g_s}}\exp\left(-\Phi^{(1)}/g_s\right).
\end{equation}

We can also work out the holomorphic limit of the 1-instanton
amplitude in a non-$A$-frame.  As argued in \cite{Gu:2022sqc} and
demonstrated below, the propagator $S$ has the property that when
replaced by its holomorphic limit $\CS$ in a non-$A$-frame with flat
coordinate $T$, its difference from $\CS_\CA$ can always be written as
\begin{equation}
  \label{eq:S-SA}
  \CS - \CS_\CA = \frac{\alpha}{\CD T \CD\CA},
\end{equation}
so that
\begin{equation}
  \mb{D} \rightarrow \alpha \pd_T.
\end{equation}
Here $\alpha$ is a frame dependent constant and in fact is the
coefficient of the dual period in the decomposition of $\CA$ in terms
of the symplectic basis
\begin{equation}
  \CA = \alpha T_D + \ldots
\end{equation}
When $\alpha\neq 0$, we can then always redefine the prepotential
$\CF_0$ so that
\begin{equation}
  \CA = \alpha \pd_T \CF_0.
\end{equation}
We find then that in this limit,
\begin{equation}
  G \rightarrow g_s^2 \alpha \pd_T \CF^{(0)},
\end{equation}
and 
\begin{equation}
  \Phi^{(1)} \rightarrow  g_s\left(\CF^{(0)}(T) - \CF^{(0)}(T-\alpha g_s)\right).
\end{equation}
The holomorphic limit of the 1-instanton amplitude $F^{(1)}$ is then
\footnote{This solution can also be obtained directly from
  \eqref{eq:CF1A} via a frame transformation
  \cite{Marino:2024yme,Marino:2024tbx}.}
\begin{equation}
  \label{eq:CF1}
  F^{(1)} \rightarrow \CF^{(1)} = \ri\sqrt{\frac{2\pi}{g_s}}\exp\left(\CF^{(0)}(T-\alpha g_s)-\CF^{(0)}(T)\right).
\end{equation}
More explicitly, when expanded in small $g_s$, we have the
trans-series form
\begin{equation}
  \label{eq:CF1}
  \CF^{(1)} = \ri\sqrt{\frac{2\pi}{g_s}}\re^{-\CA/g_s}
  \re^{\frac{1}{2}\alpha^2t+\pi\ri\alpha\beta}
  \exp \left(-\frac{g_s\alpha^3}{6} + \sum_{n=1}g_s^n
    \sum_{g=1}^{\lfloor\frac{n+1}{2}\rfloor}
    \frac{(-\alpha)^{n+2-2g}}{(n+2-2g)!}\pd_T^{n+2-2g}\CF_g(T)\right).
\end{equation}
If we evaluate the 1-instanton amplitude for the dominant instanton
sector with action $\CA_c$ in the \texttt{large radius} frame, we find
from \eqref{eq:SLR},\eqref{eq:Sc} that \eqref{eq:S-SA} is indeed true
with $\alpha =-1$, and the 1-instanton amplitude \eqref{eq:CF1} reads
\begin{equation}
  \CF^{(1)} = \ri\sqrt{\frac{2\pi}{g_s}}\exp\left(\CF^{(0)}(t+g_s)-\CF^{(0)}(t)\right).
\end{equation}
This agrees with the 1-instanton amplitude from \cite{Okuyama:2018clk}
summarized in \eqref{eq:CZ1}.

\subsection{Multi-instanton amplitudes}
\label{sc:m-inst}

In this section, we solve multi-instanton contributions to the
non-perturbative free energy.  For reasons that will become clear, it
is more convenient to consider instead the non-perturbative partition
function in an instanton sector,
\begin{equation}
  Z = Z^{(0)}+Z^{(1)}+\ldots,
\end{equation}
or rather the reduced partition function,
\begin{equation}
  Z_r = \frac{Z}{Z^{(0)}} = 1+\sum_{n\geq 1} Z_r^{(n)},
\end{equation}
which is related to the non-perturbative free energy in the same
instanton sector by
\begin{equation}
  Z_r = \exp\Bigg(\sum_{n\geq 1} F^{(n)}\Bigg).
\end{equation}
In the $A$-frame, the non-perturbative free energy reduces to a linear
combination of $\re^{-n \CA/g_s}$ with constant coefficients, and so
should the reduced partition function,
\begin{equation}
  \label{eq:ZrA}
  \CZ_{r,\CA} = \sum_{n=1}^\infty c_n \re^{-n\CA/g_s}.
\end{equation}
We will use these as boundary conditions to fix the reduced partition
function as solutions to the holomorphic anomaly equations.

Assuming that both $\log Z^{(0)}$ and $\log Z$ satisfy the holomorphic
anomaly equation \eqref{eq:hae-master}, one can derive that the
reduced partition function satisfies the following version of
holomorphic anomaly equation
\begin{equation}
  \label{eq:hae-Zr}
  \pd_SZ_r = \frac{g_s^2}{2}\CD^2Z_r + \CD \wt{F}^{(0)}\CD Z_r.
\end{equation}
It is similar to \eqref{eq:hae-F1} and is also linear and homogeneous.
This means that we can focus on the simplest boundary condition
\begin{equation}
  \label{eq:Zrn-A}
  \CZ_{r,\CA}^{(n)} = c_n \re^{-n\CA/g_s},
\end{equation}
with constant $c_n$ and look for the corresponding solution. In
addition, given the similarity with \eqref{eq:hae-F1}, we find the
same solution as \eqref{eq:F1-sol},
\begin{equation}
  \label{eq:Zrn}
  Z_{r}^{(n)} = c_n \exp(-\Phi^{(n)}/g_s),
\end{equation}
with now the exponent $n$ matching the instanton order and the
specific normalization constant $n$.  In the holomorphic limit of the
$A$-frame, it is easy to see that \eqref{eq:Zrn} indeed reduces to
\eqref{eq:Zrn-A}.  On the other hand, in the holomorphic limit of a
non-$A$-frame, the derivative $\mb{D}$ becomes
\begin{equation}
  \mb{D} = \alpha \pd_T
\end{equation}
and we find that
\begin{equation}
  \CZ_{r}^{(n)} = c_n \re^{\CF(T-n\alpha g_s)-\CF(T)} =
  \frac{1}{Z^{(0)}}c_n\re^{-n\alpha g_s\pd_T}Z^{(0)}.
\end{equation}
This is the proposition that we will use repeatedly in the following:
once its boundary condition in the $A$-frame is known, the reduced
partition function in a non-$A$-frame can be obtained by the rule of
replacelement
\begin{equation}
  \label{eq:rep}
  \re^{-\CA/g_s} \rightarrow \re^{-\alpha g_s\pd_T}.
\end{equation}

Let us now clarify what kind of reduced partition function we wish to
calculate, in other words, what exactly are the appropriate boundary
conditions, and what are the constants $c_n$ in \eqref{eq:ZrA}.
Recall that we wish to focus on the non-perturbative corrections in
the resurgent structure of the perturbative series.  In particular,
they should appear on the right hand side of the formula of Stokes
discontinuity \eqref{eq:disc} across certain Stokes ray.  It turns out
that in the case that there are multi-instanton amplitudes, instead of
Stokes discontinuity, it is more useful to consider the Stokes
transformation $\mf{S}_{\vartheta_\alpha}$ associated to the Stokes
ray $\rho_\alpha$ \cite{Ecalle}. It is an automorphism of
trans-series,
\begin{equation}
  \mf{S}_{\vartheta_\alpha}: \; \varphi(z) \rightarrow \varphi(z) + \mc{S}_\alpha
  \varphi^{(\alpha)}(z) + \ldots
\end{equation}
so that,
\begin{equation}
  \label{eq:S+S-G}
  \mr{S}^{(+)}\varphi(z) = \mr{S}^{(-)} \mf{S}_{\vartheta_\alpha} \varphi(z).
\end{equation}
As an automorphism, the Stokes transformation can be written as
exponential of differential operators known as pointed alien
derivatives \cite{Ecalle}
\begin{equation}
  \label{eq:S-Del}
  \mf{S} = \exp \left(\sum_{n=1}^\infty \alien{n\CA}\right).
\end{equation}
The alien derivative $\alien{n\CA}$ serves to send a trans-series of
the order $\re^{-\ell\CA/g_s}$ to another trans-series of the order
$\re^{-(\ell+n)\CA/g_s}$, and it behaves truely like a derivative,
satisfying the Leibniz rule and the chain rule.  In terms of alien
derivatives, the action of the Stokes transformation reads
\begin{align}
  \mf{S}\varphi(z) = \varphi(z) + \alien{\CA}\varphi(z) +
  \left(\alien{2\CA}+\frac{1}{2}\alienpwr{\CA}{2}\right)\varphi(z)+\ldots
\end{align}
This formula tells us that when there are multi-instantons, their
amplitudes are in general not unique\footnote{This has also been
  emphasized in the trans-series formalism of eigen-energies in
  quantum mechanics, where there are $n$ instanton amplitudes at the
  $n$-instanton order.  See for instance
  \cite{vanSpaendonck:2023znn}.}.  At the order of $\re^{-2\CA/g_s}$,
there are two instanton amplitudes
\begin{equation}
  \alien{2\CA}\varphi(z), \;\; \alienpwr{\CA}{2}\varphi(z),
\end{equation}
and at the order of $\re^{-3\CA/g_s}$, there are three instanton amplitudes
\begin{equation}
  \alien{3\CA}\varphi(z), \;\; \alien{2\CA}\alien{\CA}\varphi(z), \;\; \alienpwr{\CA}{3}\varphi(z),
\end{equation}
and etc.  Only in the special case where $\alien{n\CA}\varphi(z)$
truncate to polynomials of finite degrees so that further application
of alien derivatives vanishes do we have a single instanton amplitude
at each order.

In the case of topological string free energy, we will first work out
the basic $n$-instanton amplitude $F^{(n)}$ defined by,
\begin{equation}
  \alien{n\CA} F^{(0)} = F^{(n)}.
\end{equation}
Here we have absorbed the Stokes constant inside $F^{(n)}$ or simply
set it to one.  All the other $n$-instanton amplitudes can be derived
from the basic instanton amplitudes as we will see later.

We first work out the boundary conditions, i.e. the instanton
amplitudes in the $A$-frame.  On the one hand, we know from
\eqref{eq:CFnA} that
\begin{equation}
  \mf{S}\CF^{(0)}_\CA = \CF^{(0)} + \sum_{n=1}^\infty
  \mu_n \re^{-n\CA/g_s}.
\end{equation}
On the other hand, \eqref{eq:S-Del} implies
\begin{equation}
  \mf{S}\CF^{(0)}_\CA = \CF^{(0)} + \alien{\CA}\CF^{(0)}_\CA +
  \left(\alien{2\CA}+\frac{1}{2}\alienpwr{\CA}{2}\right)\CF^{(0)}_\CA+\ldots
\end{equation}
Given that the alien derivatives annihilate truncated power series, we
find recursively that
\begin{equation}
  \alien{n\CA}\CF^{(0)}_\CA = \CF^{(n)}_\CA = \mu_n \re^{-n\CA/g_s}.
\end{equation}

To find the basic instanton amplitude $F^{(n)}$ at order $n$, we
single out the contribution of $\alien{n\CA}$ and ignore for the
moment all the other alien derivatives.
We choose the special boundary condition that
\begin{equation}
  \Gamma_n:\quad \alien{\ell\CA}\CF^{(0)}_{\CA} =
  \delta_{n,\ell} \CF^{(n)}_\CA,
\end{equation}
so that
\begin{equation}
  \Gamma_n:\quad \mf{S}\CF^{(0)}_{\CA}  =
  \CF^{(0)}_{\CA} + \CF^{(n)}_\CA.
\end{equation}
We use the label $\Gamma_n$ to emphasize the special boundary
condition that we have chosen.  The partition function then reads
\begin{equation}
  \Gamma_n:\quad \mf{S}\CZ^{(0)}_{\CA} = \CZ_{\CA}^{(0)}\exp \CF_\CA^{(n)},
\end{equation}
and the reduced partition function after crossing the Stokes ray is
\begin{equation}
  \Gamma_n:\quad\CZ_{r,\CA} = \exp \CF^{(n)}_\CA = \sum_{\ell=0}^\infty
  \frac{1}{\ell!}\mu_n^\ell \re^{-n\ell \CA/g_s}.
\end{equation}
By our replacement rule \eqref{eq:rep}, the reduced partition function
in a non-$A$-frame is
\begin{equation}
  \Gamma_n:\quad\CZ_{r} = \sum_{\ell=0}^\infty
  \frac{1}{\ell!}\mu_n^\ell \re^{\CF^{(0)}(T-n\ell\alpha g_s)-\CF^{(0)}(T)},
\end{equation}
and the non-perturbative free energy is
\begin{equation}
  \Gamma_n:\quad\CF = \log \CZ= \CF^{(0)} + \mu_n
  \re^{\CF^{(0)}(T-n\alpha g_s)-\CF^{(0)}(T)} + \CO(\re^{-2n\CA/g_s}).
\end{equation}
By comparing with
\begin{equation}
  \Gamma_n:\quad\CF = \mf{S}\CF^{(0)} =
  \left(\exp\alien{n\CA}\right)\CF^{(0)} = \CF^{(0)} +
  \alien{n\CA}\CF^{(0)} +\CO(\re^{-2n\CA/g_s}),
\end{equation}
we conclude that the basic instanton amplitude at order $n$ is
\begin{equation}
  \label{eq:Fn}
  \CF^{(n)} = \alien{n\CA} \CF^{(0)} =
  \mu_n \re^{\CF^{(0)}(T-n\alpha g_s)-\CF^{(0)}(T)}.
\end{equation}

An important feature of the basic instanton amplitude is the shift by
$n$ units of the flat coordinate, as emphasized in \cite{Gu:2022sqc},
and it corresponds to $n$ units of the RR-flux through the compact
$T^2$.  In addition, it is expressed as a functional of the
perturbative free energy.  Together with the chain rule of the alien
derivative, it allows us to calculate the other instanton amplitudes
at multi-instanton orders.  For instance, at $2$-instanton order, the
other instanton amplitude is
\begin{equation}
  \label{eq:alien-2A}
  \alienpwr{\CA}{2}\CF^{(0)} =
  \mu_1^2\left(\re^{\CF_2^{(0)}-\CF^{(0)}}-\re^{2\CF_1^{(0)}-2\CF^{(0)}}\right).
\end{equation}
At 3-instanton order,
\begin{align}
  \alien{\CA}\alien{2\CA}\CF^{(0)} =
  &\mu_1\mu_2
    \left(\re^{\CF^{(0)}_3-\CF^{(0)}}-\re^{\CF_2^{(0)}+\CF_1^{(0)}-2\CF^{(0)}}\right),
  \label{eq:alien-3A}\\
  \alienpwr{\CA}{3}\CF^{(0)} =
  &\mu_1^3
    \left(\re^{\CF_3^{(0)}-\CF^{(0)}}-3\re^{\CF_2^{(0)}+\CF_1^{(0)}-2\CF^{(0)}}
    +2\re^{3\CF_1^{(0)}-3\CF^{(0)}}\right).
    \label{eq:alien-3B}
\end{align}

We now check our results explicitly.  In terms of alien derivatives
and up to 4-instanton order, the Stokes discontinuity is
\begin{align}
  \text{disc}_0\CF^{(0)} =
  &\left(\exp\sum_{n=1}^\infty \alien{n\CA}\right)\CF^{(0)} -\CF^{(0)} \nn=
  &\alien{\CA}\CF^{(0)} +
    \left(\alien{2\CA}+\frac{1}{2}\alienpwr{\CA}{2}\right)\CF^{(0)}\nn+
  &\left(\alien{3\CA}+\alien{2\CA}\alien{\CA}+\frac{1}{3!}\alienpwr{\CA}{3}\right)\CF^{(0)}
    \nn +
  &\left(\alien{4\CA}+\alien{3\CA}\alien{\CA}+\frac{1}{2}\alienpwr{2\CA}{2}
    +\frac{1}{2}\alien{2\CA}\alienpwr{\CA}{2}+\frac{1}{4!}\alienpwr{\CA}{4}\right)
    \CF^{(0)} + \ldots
    \label{eq:disc0-alien}
\end{align}
We can check \eqref{eq:disc0-alien} together with \eqref{eq:Fn},
\eqref{eq:alien-2A}, \eqref{eq:alien-3A}, \eqref{eq:alien-3B}
numerically.
As we have seen in Figs.~\ref{fig:brl} and also from
Section~\ref{sc:bdy}, when $t$ is positive, there is a Stokes ray
along the positive real axis due to the Borel singularities
$\ell \CA_c$ with $\ell=1,2,3,\ldots$.  We take $t = 2\pi/5$,
$t=2\pi/7$, calculate numerically the Stokes discontinuity across the
positive real axis at $g_s = \CA_c/10$, and subtract successively
instanton amplitudes of higher and higher orders.  We find that the
remainder becomes progressively smaller as shown in
Tab.~\ref{tab:alien}.
We also note that our results only depend on the boundary conditions
\eqref{eq:CFnA} and the HAEs \eqref{eq:hae}, and should in fact be
valid even when $\theta$ is turned on so that $t$ is complex as well.

\begin{table}
  \centering
  \begin{tabular}{*{3}{>{$}c<{$}}}\toprule
    n
    & t=2\pi/5
    & t=2\pi/7 \\
    \midrule
    0
    & -\ri\,0.002031
    & -\ri\,0.0003390 \\
    1
    & +1.5160\times 10^{-8} + \ri\,4.3236\times 10^{-10}
    & +3.4822\times 10^{-8} + \ri\,1.2783\times 10^{-9}\\
    2
    & -8.2065\times 10^{-14} + \ri\,1.7062\times 10^{-12}
    & -3.6852\times 10^{-13} + \ri\,5.6901\times 10^{-12}\\
    3
    & -2.1957\times 10^{-16} - \ri\,1.4307\times 10^{-17}
    & -1.0821\times 10^{-15} - \ri\,9.6552\times 10^{-17}\\
    4
    & +2.4669\times 10^{-21} - \ri\,3.0214\times 10^{-20}
    & +2.4866\times 10^{-20} - \ri\,2.2110\times 10^{-19}\\
    5
    & +4.3291\times 10^{-24} - \ri\,1.0811\times 10^{-23}
    & +4.7115\times 10^{-23} + \ri\,9.5123\times 10^{-24}\\
    \bottomrule
  \end{tabular}
  \caption{Stokes discontinuity of perturbative free energy across the
    positive real axis after subtracting up to order $n$ instanton
    amplitudes at $g_s = \CA_c/10$. We use perturbative power series
    up to 200 terms, and the expected error due to the truncation is
    of the order $\CO(10^{-24})$ in both cases.}
  \label{tab:alien}
\end{table}


We can also work out the Stokes transformation of the partition
function.
In the $A$-frame, the partition function after the Stokes
transformation is
\begin{equation}
  \mf{S}\CZ^{(0)}_\CA = \exp \mf{S}\CF^{(0)}_\CA = \CZ^{(0)}_\CA
  \exp\sum_{n=1}^\infty \mu_n \re^{-n\CA/g_s},
\end{equation}
and the reduced partition function is
\begin{equation}
  \CZ_{r,\CA} = \mf{S}\CZ^{(0)}_\CA/\CZ^{(0)}_\CA = \exp\sum_{n=1}^\infty \mu_n \re^{-n\CA/g_s}.
\end{equation}
With the Fa\`a di Bruno formula
\begin{equation}
  \label{eq:FdB-exp}
  \exp\left(\sum_{j=1}^\infty x_j \frac{t^j}{j!}\right) = \sum_{n=0}^\infty B_n(\{x_j\}_{j=1,\ldots,n})\frac{t^n}{n!},
\end{equation}
where $B_n(\ldots)$ are the Bell polynomials, the reduced partition
function in the $A$-frame can be put explicitly in the form
\eqref{eq:ZrA} that we anticipate
\begin{equation}
  \CZ_{r,\CA} =\sum_{n=0}^\infty \frac{1}{n!}
  B_n\left(\{j!\mu_j\}_{j=1,\ldots,n}\right)\re^{-n\CA/g_s}.
\end{equation}
Apply again the rule of replacement \eqref{eq:rep}, the reduced
partition function in a non-$A$-frame reads
\begin{equation}
  \CZ_r = \sum_{n=0}^\infty
  \frac{1}{n!}B_n\left(\{j!\mu_j\}_{j=1,\ldots,n}\right)
  \re^{\CF(T-n\alpha g_s) - \CF^{(0)}(T)},
\end{equation}
and the non-perturbative partition function after the Stokes
transformation is
\begin{equation}
  \label{eq:SZ}
  \mf{S}\CZ^{(0)} = \sum_{n=0}^\infty
  \frac{1}{n!}B_n\left(\{j!\mu_j\}_{j=1,\ldots,n}\right)
  \CZ^{(0)}(T-n\alpha  g_s) 
  .
\end{equation}

The result \eqref{eq:SZ} also allows us to write down a compact
formula for the right hand side of the Stokes discontinuity formula
\eqref{eq:disc0-alien}.
With the Fa\`a di Bruno formula for a second time
\begin{equation}
    \label{eq:FdB-log}
    \log\left(1+\sum_{j=1}^\infty\frac{x_j}{j!}t^j\right) =
    \sum_{n=1}^\infty \frac{t^n}{n!}\sum_{k=1}^n(-1)^{k-1}(k-1)!
    B_{n,k}(\{x_j\}_{j=1,\ldots,n-k+1}),
\end{equation}
where $B_{n,k}(\ldots)$ are the incomplete Bell polynomials,
we find that
\begin{align}
  &\text{disc}_0\CF^{(0)}  =
    \mf{S}\CF^{(0)} - \CF^{(0)} = \log\mf{S}\CZ^{(0)} -
    \log\CZ^{(0)}\\  =
  &\sum_{n=1}^\infty\frac{1}{n!}\sum_{k=1}^n(-1)^{k-1}(k-1)!\re^{-k\CF^{(0)}}
    B_{n,k}\left(\{
    B_\ell(\{j!\mu_j\}_{j=1,\ldots,\ell})\re^{\CF^{(0)}(T-\ell\alpha g_s)}
    \}_{\ell=1,\ldots,n-k+1}\right),\nonumber
\end{align}
from which effects of alien derivatives such as \eqref{eq:alien-2A},
\eqref{eq:alien-3A}, \eqref{eq:alien-3B} and those of even higher
orders can be easily read off.

\subsection{Non-perturbative proposal}
\label{sc:med}

Consider the holomorphic partition function in the \texttt{large
  radius} frame.  When both $t$ and $g_s$ are positive, the Borel
resummed partition function has a Stokes ray along the positive real
axis induced by real instantons with action $\ell t^2/2$
($\ell=1,2,\ldots$), and the Stokes transformation across this Stokes
ray is described by \eqref{eq:SZ} with $T =t,\alpha=-1$.  This tells
us how to write down the non-perturbative partition function which
includes the contributions of all the real instantons, and which is
itself real for positve $t$ and $g_s$.  It is given by
\begin{equation}
  \label{eq:Znps}
  \CZ^{\text{np}}(t;g_s;\boldsymbol{\sigma}) =
  \sum_{n=0}^\infty\frac{1}{n!}
  B_n\left(\left\{\sigma_{j}\pm\frac{\ri}{2}j!
      \sqrt{\frac{2\pi}{g_s}}\frac{(-1)^{j }}
      {j^{3/2}}\right\}_{j=1,\ldots,n}\right)
  \mr{S}^{(\pm)}\CZ(t+ng_s;g_s),
\end{equation}
where $\boldsymbol{\sigma}=(\sigma_1,\sigma_2,\ldots)$ is an infinite
family of \emph{real} parameters.  Here we can take either of the two
lateral Borel resummations $\mr{S}^{(\pm)}$, but at the same time we
have to choose the same sign $\pm$ inside the Bell polynomials to
match it.  The two definitions are equivalent due to the Stokes
transformation \eqref{eq:SZ} as
\begin{align}
  &\sum_{n=0}^\infty\frac{1}{n!}
    B_n\left(\left\{\sigma_{j}+\frac{\ri}{2}j!
    \sqrt{\frac{2\pi}{g_s}}\frac{(-1)^{j }}
    {j^{3/2}}\right\}_{j=1,\ldots,n}\right)
    \mr{S}^{(+)}\CZ(t+ng_s;g_s) \nn=
  &\sum_{n=0}^\infty\frac{1}{n!}
    B_n\left(\left\{\sigma_{j}+\frac{\ri}{2}j!
    \sqrt{\frac{2\pi}{g_s}}\frac{(-1)^{j }}
    {j^{3/2}}\right\}_{j=1,\ldots,n}\right)
    \mr{S}^{(-)}\mf{S}\CZ(t+ng_s;g_s) \nn=
  &\sum_{n=0}^\infty\frac{1}{n!}\mr{S}^{(-)}\CZ(t+ng_s;g_s)
    \sum_{m=0}^n{n\choose m}B_m\left(\left\{j!\,\mu_j\right\}_{j=1,\ldots,m}\right)\nn
  &\times B_{n-m}\left(\left\{\sigma_{j}+\frac{\ri}{2}j!
    \sqrt{\frac{2\pi}{g_s}}\frac{(-1)^{j }}
    {j^{3/2}}\right\}_{j=1,\ldots,n-m}\right)\nn=
  &\sum_{n=0}^\infty\frac{1}{n!}
    B_n\left(\left\{\sigma_{j}-\frac{\ri}{2}j!
    \sqrt{\frac{2\pi}{g_s}}\frac{(-1)^{j }}
    {j^{3/2}}\right\}_{j=1,\ldots,n}\right)
    \mr{S}^{(-)}\CZ(t+ng_s;g_s),
\end{align}
where $\mu_j$ are defined in \eqref{eq:CFnA} and we have used the law
of convolution of Bell polynomials 
\begin{equation}
  \label{eq:Bcon}
  \sum_{m=0}^n {n \choose m}B_{m}(\{x\})B_{n-m}(\{y\}) = B_n(\{x+y\}).
\end{equation}
This is also verified numerically in Tabs.~\ref{tab:Znp5},
\ref{tab:Znp3}, \ref{tab:Znp5s}, \ref{tab:Znp3s}: as higher order
instanton corrections are included, both of the two lateral
resummations have progressively smaller imaginary parts, and they
converge to the same real number.

\jjj{Modification begins} We note that the formula for
non-perturbative partition function \eqref{eq:Znps} is \emph{not
  unique} as it depends on an infinite family of free parameters
$\bs\sigma$.  It thus seems that the resurgence analysis does not
capure the complete information of non-perturbative corrections.  We
make two comments here.  Firstly, it is not uncommon that the
resurgence analysis fails to uncover the full non-perturbative
corrections.  For instance, in quantum mechanical models whose
potentials are symmetric, resurgence analysis does not capture
non-perturbative information in the energy spectrum that arise from
(quantum mechanically breaking) these symmetries, such as the removal
of energy degeneracy in the double-well model with parity symmetry,
and the Bloch angle in the cosine model with periodic symmetry (see
for instance discussion in \cite{Marino2015:instantons}).  Second,
nevertheless, there does not seem to be any additional symmetries in
2d YM which are quantum mechanically broken, parameterized by some
order parameter. Therefore, we conjecture that we can simply take the
simplest member in the family \eqref{eq:Znps}, which is the celebrated
medium resummation of the perturbative partition function
\jjj{Modification ends.}
\begin{align}
  \label{eq:Zmed}
  \CZ^{\text{np}}(t;g_s;\boldsymbol{0}) =
  &\mr{S}^{\text{med}}\CZ(t;g_s) = \mr{S}^{(\pm)}\mf{S}^{\mp 1/2}
    \CZ(t;g_s) \\=
  &\sum_{n=0}^\infty\frac{1}{n!}B_n\left(\left\{\pm\frac{\ri}{2}j!
    \sqrt{\frac{2\pi}{g_s}}\frac{(-1)^{j
    }}{j^{3/2}}\right\}_{j=1,\ldots,n}\right)
    \mr{S}^{(\pm)}\CZ(t+ng_s;g_s).\nonumber    
\end{align}


\begin{table}
  \centering
  \begin{tabular}{*{3}{>{$}c<{$}}}\toprule
    & \CZ^{\text{np}} \text{ by } + & \CZ^{\text{np}} \text{ by } -\\\midrule
    n=0
    & 2.95\times10^{-6}+\ri\,2.95\times 10^{-8}& 2.95\times10^{-6}-\ri\,2.95\times 10^{-8}\\
    n=1
    & 2.95\times10^{-6}-\ri\,8.26\times 10^{-12}& 2.95\times10^{-6}+\ri\,8.26\times 10^{-12}\\
    n=2
    & 2.95\times10^{-6}-\ri\,5.76\times 10^{-15}& 2.95\times10^{-6}+\ri\,5.76\times 10^{-15}\\
    n=3
    & 2.95\times10^{-6}+\ri\,1.94\times 10^{-19}& 2.95\times10^{-6}-\ri\,1.94\times 10^{-19}\\
    n=4
    & 2.95\times10^{-6}+\ri\,3.26\times 10^{-24}& 2.95\times10^{-6}-\ri\,3.26\times 10^{-24}\\
    \bottomrule
  \end{tabular}
  \caption{The non-perturbative partition function
    $\CZ^{\text{np}}(t;g_s;\boldsymbol{0})$ at $t = 2\pi/5$,
    $g_s = \CA_c/5$ including up to order $n$ instanton corrections.
    We use perturbative series up to $200$ terms, and the expected
    error due to the truncation is of the order $\CO(10^{-24})$.}
  \label{tab:Znp5}
\end{table}

\begin{table}
  \centering
  \begin{tabular}{*{3}{>{$}c<{$}}}\toprule
    & \CZ^{\text{np}} \text{ by } + & \CZ^{\text{np}} \text{ by } -\\\midrule
    n=0
    & 4.48\times10^{-4}+\ri\,1.77\times 10^{-6} & 4.48\times10^{-4}-\ri\,1.77\times 10^{-6}\\
    n=1
    & 4.48\times10^{-4}-\ri\,1.08\times 10^{-10}& 4.48\times10^{-4}+\ri\,1.08\times 10^{-10}\\
    n=2
    & 4.48\times10^{-4}-\ri\,1.10\times 10^{-15}& 4.48\times10^{-4}+\ri\,1.10\times 10^{-15}\\
    n=3
    & 4.48\times10^{-4}+\ri\,2.47\times 10^{-19}& 4.48\times10^{-4}-\ri\,2.47\times 10^{-19}\\
    \bottomrule
  \end{tabular}
  \caption{The non-perturbative partition function
    $\CZ^{\text{np}}(t;g_s;\boldsymbol{0})$ at $t = 2\pi/3$,
    $g_s = \CA_c/5$ including up to order $n$ instanton corrections.
    We use perturbative series up to $200$ terms, and the expected
    error due to the truncation is of the order $\CO(10^{-20})$.}
  \label{tab:Znp3}
\end{table}

\begin{table}
  \centering
  \begin{tabular}{*{3}{>{$}c<{$}}}\toprule
    & \CZ^{\text{np}} \text{ by } + & \CZ^{\text{np}} \text{ by } -\\\midrule
    n=0
    & 2.95\times10^{-6}+\ri\,2.95\times 10^{-8} & 2.95\times10^{-6}-\ri\,2.95\times 10^{-8}\\
    n=1
    & 2.98\times10^{-6}+\ri\,6.20\times 10^{-12} & 2.98\times10^{-6}-\ri\,6.20\times 10^{-12}\\
    n=2
    & 2.98\times10^{-6}+\ri\,1.17\times 10^{-15}& 2.98\times10^{-6}-\ri\,1.17\times 10^{-15}\\
    n=3
    & 2.98\times10^{-6}+\ri\,3.51\times 10^{-19}& 2.98\times10^{-6}-\ri\,3.51\times 10^{-19}\\
    n=4
    & 2.98\times10^{-6}+\ri\,5.11\times 10^{-24}& 2.98\times10^{-6}-\ri\,5.11\times 10^{-24}\\
    \bottomrule
  \end{tabular}
  \caption{The non-perturbative partition function
    $\CZ^{\text{np}}(t;g_s,\boldsymbol{\sigma})$ at $t = 2\pi/5$,
    $g_s = \CA_c/5$ with $\boldsymbol{\sigma}=(3,2,1,9,\ldots)$
    including up to order $n$ instanton corrections.  We use
    perturbative series up to $200$ terms, and the expected error due
    to the truncation is of the order $\CO(10^{-24})$.}
  \label{tab:Znp5s}
\end{table}

\begin{table}
  \centering
  \begin{tabular}{*{3}{>{$}c<{$}}}\toprule
    & \CZ^{\text{np}} \text{ by } + & \CZ^{\text{np}} \text{ by } -\\\midrule
    n=0
    & 4.48\times10^{-4}+\ri\,1.77\times 10^{-6} & 4.48\times10^{-4}-\ri\,1.77\times 10^{-6}\\
    n=1
    & 4.52\times10^{-4}+\ri\,1.41\times 10^{-9} & 4.48\times10^{-4}+\ri\,1.41\times 10^{-9}\\
    n=2
    & 4.52\times10^{-4}+\ri\,3.21\times 10^{-14}& 4.48\times10^{-4}+\ri\,2.21\times 10^{-15}\\
    n=3
    & 4.52\times10^{-4}+\ri\,2.65\times 10^{-19}& 4.48\times10^{-4}-\ri\,2.65\times 10^{-19}\\
    \bottomrule
  \end{tabular}
  \caption{The non-perturbative partition function
    $\CZ^{\text{np}}(t;g_s;\boldsymbol{0})$ at $t = 2\pi/3$,
    $g_s = \CA_c/5$ with $\boldsymbol{\sigma}=(5,3,7,1,\ldots)$
    including up to order $n$ instanton corrections.  We use
    perturbative series up to $200$ terms, and the expected error due
    to the truncation is of the order $\CO(10^{-20})$.}
  \label{tab:Znp3s}
\end{table}

Let us compare this result with the prosposal of Okuyama and Sakai
\cite{Okuyama:2018clk} summarized in Section~\ref{sc:os}.  We can
spell out the latter more explicitly.  By using the Fa\`a di Bruno
formula \eqref{eq:FdB-exp}, we find
\begin{equation}
  \phi_n(q) = \frac{1}{n!}B_n\left(\left\{\frac{\ri}{2}(j-1)!
      (-1)^j \vartheta_2(q^j)\right\}_{j=1,\ldots,n}\right).
\end{equation}
In the limit of $g_s\rightarrow 0$ and $q\rightarrow 1$, by the
modular property of Jacobi theta functions
\begin{equation}
  \vartheta_2(q^\ell) = \vartheta_2\left(0;\frac{\ri\ell
      g_s}{2\pi}\right) = \sqrt{\frac{2\pi}{\ell g_s}}
  \vartheta_4\left(0;\frac{2\pi\ri}{\ell g_s}\right) =
  \sqrt{\frac{2\pi}{\ell g_s}}
  \vartheta_4\left(\re^{-4\pi^2/(\ell g_s)}\right),
\end{equation}
\eqref{eq:ZOS-phi} can be written as
\begin{equation}
  \label{eq:S+ZOS}
  Z^{\text{OS}}(t;g_s) = \sum_{n=0}^\infty
  \frac{1}{n!}B_n\left(\left\{
      \frac{\ri}{2} j!
      \sqrt{\frac{2\pi}{g_s}}\frac{(-1)^j}{j^{3/2}}\vartheta_4(\re^{-4\pi^2/(j
        g_s)})
    \right\}\right) \mr{S}^{(+)}Z^{\text{top}}(t+ng_s;g_s).
\end{equation} 

Eq.~\eqref{eq:S+ZOS} indicates that there are instanton sectors with
action $nt^2/2 + 4 m \pi^2/j$, with $j=1,\ldots,n$ and $m=1,2,\ldots$.
They are not observed in the numerical calculations of alien
derivatives.  This problem can be bypassed by arguing that these
additional instanton sectors are in disjoint resurgent structures than
that of the perturbative partition function, therefore inaccessible by
resurgence analysis of the perturbative sector, just like even and odd
instanton sectors in the double-well quantum mechanics model.

More importantly, \eqref{eq:S+ZOS} coincides with \eqref{eq:Zmed} with
$+$ sign in the leading order with $\vartheta_4\approx 1$, but
deviates from it by higher order terms.  This indicates that
\eqref{eq:S+ZOS} is \emph{not} real for positive $t$ and $g_s$, i.e.~the
cancellation of imaginary part is incomplete due to higher order terms
in $\vartheta_4$.
To see this problem more clearly, 
let us apply the Stokes transformation \eqref{eq:S+S-G},\eqref{eq:SZ},
\begin{align}
  \CZ^{\text{OS}}(t;g_s) =
  &\sum_{n=0}^\infty
    \frac{1}{n!}B_n\left(\left\{
    \frac{\ri}{2} j!
    \sqrt{\frac{2\pi}{g_s}}\frac{(-1)^j}{j^{3/2}}
    \vartheta_4(\re^{-4\pi^2/(j g_s)})
    \right\}\right)\mr{S}^{(-)}\mf{S}Z(t+ng_s;g_s)\nn=
  &\sum_{n=0}^\infty\frac{1}{n!} \mr{S}^{(-)}Z(t+ng_s;g_s)
    \sum_{m=0}^n{n \choose m} B_m\left(\left\{ -\ri j!
    \sqrt{\frac{2\pi}{g_s}}\frac{(-1)^j}{j^{3/2}}    
    \right\}\right)\nn   
  &\phantom{===}\times B_{n-m}\left(\left\{
    \frac{\ri}{2} j!
    \sqrt{\frac{2\pi}{g_s}}\frac{(-1)^j}{j^{3/2}}
    \vartheta_4(\re^{-4\pi^2/(j g_s)})
    \right\}\right)\nn=
  &\sum_{n=0}^\infty\frac{1}{n!} 
    B_{n}\left(\left\{
    -\frac{\ri}{2} j!
    \sqrt{\frac{2\pi}{g_s}}\frac{(-1)^j}{j^{3/2}}
    (2-\vartheta_4(\re^{-4\pi^2/(j g_s)}))
    \right\}\right)\mr{S}^{(-)}\CZ(t+ng_s;g_s),
    \label{eq:S-ZOS}
\end{align}
where we have used again \eqref{eq:Bcon}.  Note that the right hand
side of \eqref{eq:S-ZOS} and \eqref{eq:S+ZOS} are not symmetric as in
the case of \eqref{eq:Zmed}, which implies that the Borel resummation
\eqref{eq:S+ZOS} is \emph{not} real.

\begin{table}
  \centering
  \begin{tabular}{*{3}{>{$}c<{$}}}\toprule
    & \CZ^{\text{np}} \text{ by } + & \CZ^{\text{OS}}\\\midrule
    n=0
    & 3.21\times10^{-5}+\ri\,4.38\times 10^{-11}& 3.21\times10^{-5}-\ri\,4.38\times 10^{-11}\\
    n=1
    & 3.21\times10^{-5}-\ri\,6.07\times 10^{-21}& 3.21\times10^{-5}+\ri\,3.98\times 10^{-15}\\
    n=2
    & 3.21\times10^{-5}+\ri\,9.40\times 10^{-23}& 3.21\times10^{-5}+\ri\,3.98\times 10^{-15}\\
    n=3
    & 3.21\times10^{-5}+\ri\,9.40\times 10^{-23}& 3.21\times10^{-5}+\ri\,3.98\times 10^{-15}\\
    \bottomrule
  \end{tabular}
  \caption{The non-perturbative partition function
    $\CZ^{\text{np}}(t;g_s;\boldsymbol{0})$ and the non-perturbative
    partition function $\CZ^{\text{OS}}$ proposed by Okuyama-Sakai, at
    $t = 2\pi$, $g_s = \CA_c/10$, including up to order $n$ instanton
    corrections.  We use perturbative series up to $200$ terms, and
    the expected error due to truncation is of the order
    $\CO(10^{-24})$.}
  \label{tab:ZOS1}
\end{table}

\begin{table}
  \centering
  \begin{tabular}{*{3}{>{$}c<{$}}}\toprule
    & \CZ^{\text{np}}\text{ by } + & \CZ^{\text{OS}}\\\midrule
    n=0
    & 1.48\times10^{-7}+\ri\,9.83\times 10^{-13}& 1.48\times10^{-7}+\ri\,9.84\times 10^{-13}\\
    n=1
    & 1.48\times10^{-7}-\ri\,1.09\times 10^{-20}& 1.48\times10^{-7}-\ri\,1.06\times 10^{-20}\\
    n=2
    & 1.48\times10^{-7}-\ri\,3.17\times 10^{-27}& 1.48\times10^{-7}+\ri\,3.33\times 10^{-22}\\
    n=3
    & 1.48\times10^{-7}-\ri\,3.17\times 10^{-27}& 1.48\times10^{-7}+\ri\,3.33\times 10^{-22}\\
    \bottomrule
  \end{tabular}
  \caption{The non-perturbative partition function
    $\CZ^{\text{np}}(t;g_s;\boldsymbol{0})$ and the non-perturbative
    partition function $\CZ^{\text{OS}}(t;g_s)$ proposed by
    Okuyama-Sakai, at $t = 4\pi/3$, $g_s = \CA_c/10$, including up to
    order $n$ instanton corrections.  We use perturbative series up to
    $200$ terms, and the expected error due to truncation is of the
    order $\CO(10^{-27})$.}
  \label{tab:ZOS2}
\end{table}

We verify this numerically in Tabs.~\ref{tab:ZOS1}, \ref{tab:ZOS2}: as
the order of included instanton corrections increases, the imaginary
part of the resummed partition function \eqref{eq:S+ZOS} stops
decreasing at some point, even though it is still much larger than the
expected error due to the truncation of the perturbative power
series\footnote{This is probably the reason that in
  \cite[Sec.~5]{Okuyama:2018clk} the two sides of \eqref{eq:ZOS-phi}
  differ in numerical tests by a small imaginary part when $N$ is
  large.}, in stark contrast to the medium resummed non-perturbative
partition function \eqref{eq:Zmed}.\footnote{To see the effect of
  $\vartheta_4(\re^{-4\pi^2/(jg_s)})$, one has to choose $t$
  appropriately so that the effect of $\re^{-2\pi^2/(jg_s)}$ is
  comparable with instanton corrections of the order
  $\re^{-t^2/(2g_s)} = \re^{2\pi^2\tau^2/g_s}$.}

\subsection{Stokes spectrum}
\label{sc:BPS}

In the previous sections, we have focused on the instanton sectors
with action $\pm n\CA_c$ with $n=1,2,3,\ldots$.  As showcased in
Figs.~\ref{fig:brl}, there are additional instanton sectors with
different actions which manifest themselves as additional Borel
singularities of the perturbative free energy.  We will study their
properties, especially the Stokes transformation of the perturbative
free energy induced by them.

\begin{figure}
  \centering
  \subfloat[$t=16\pi/3$]{\includegraphics[height=4cm]{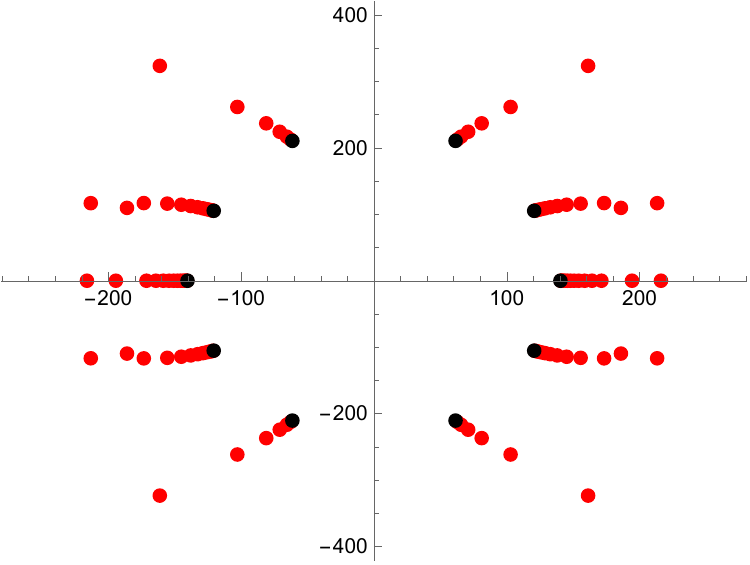}}\hspace{4ex}
  \subfloat[$t=4\pi/3\,\re^{\pi\ri/6}$]{\includegraphics[height=4cm]{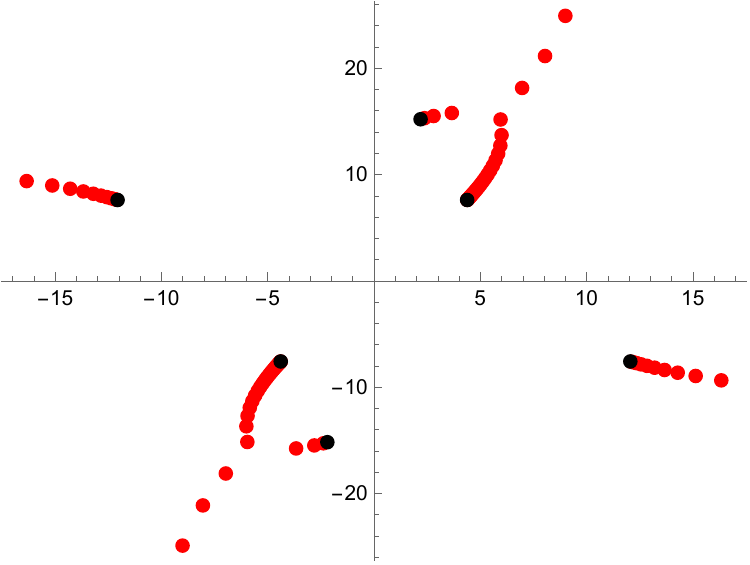}}
  \caption{Borel singularities of perturbative free energy in the
    \texttt{large radius} frame.  We use perturbative series truncated
    to 200 terms, and use Pad\'e approximant to mimic the analytic
    continuation of the Borel transform.  The singular points (red) of
    the approximation would condense to branch cuts if the truncation
    is pushed to infinity.  At $t = 16\pi/3$ $(a)$, the branch points
    (black) have charges $\pm(1,0,0)$, $\pm(1,2,2)$, $\pm(1,-2,2)$,
    $\pm(1,4,8)$, $\pm(1,-4,8)$.  At $t=4\pi/3\,\re^{\pi\ri/6}$ $(b)$,
    the branch points (black) in the 1st and 3rd quadrants have
    charges $\pm(1,0,0)$, $\pm(1,2,2)$, and the branch points in the
    2nd and 4th quadrants have charges $\pm(2,2,1)$.}
  \label{fig:brl2}
\end{figure}

Additional instanton sectors can be uncovered from the Borel
singularities of perturbative free energy in two different ways.  We
can either increase the value of $t$ or give $t$ a non-trivial
$\theta$ angle, as in the left and right panels of
Fig.~\ref{fig:brl2}.  In the former case, we find two towers of Borel
singularities located at $\CA_\gamma$ with charge vectors
\begin{equation}
  \gamma_{(m,\pm,\pm)} = \pm (1,\pm 2m,2m^2), \quad m=1,2,\ldots.
\end{equation}
We find that in the \texttt{large radius} frame, the Stokes
transformation is
\begin{equation}
  \alien{\CA_{\gamma_{(m,\pm,\pm)}}}\CF^{(0)} = g_s^2\CF^{(1)},
\end{equation}
with $\CF^{(1)}$ given explicitly by \eqref{eq:CF1} with
$\alpha=\pm 1,\beta = \pm m$, slightly modified by the additional
factor of $g_s^2$.  In the latter case, we find further Borel
singularities, such as the ones with charges
\begin{equation}
  \gamma_{\pm 2} = \pm (2,2,1).
\end{equation}
In the \texttt{large radius} frame, the Stokes transformation is
\begin{equation}
  \alien{\CA_{\gamma_{\pm 2}}}\CF^{(0)} = g_s^2\mu_2 \exp\left(\CF^{(0)}(t+2g_s)-\CF^{(0)}(t)\right),
\end{equation}
which is a bit peculiar, as it is a mixture of the 1-instanton
amplitude with $\alpha=2$ and the 2-instanton prefactor $\mu_2$.  In
all these examples, the Stokes constants are identically set to one
due to our normalization of instanton amplitudes.

Following the conjecture in \cite{Gu:2021ize,Gu:2022sqc,Gu:2023mgf},
these instanton sectors might correspond to different BPS states --
D-brane bound states -- in the type IIA string theory compactified on
the Calabi-Yau threefold $X_E$, with $\gamma$ as the charge vectors.
The BPS spectrum of type IIA string in general have wall-crossing
phenomenon in the moduli space: across a real codimension one wall of
marginal stability characterized by that two integral periods align in
the complex plane, the spectrum of BPS states changes.  Here the
condition of the wall of marginal stability translates to
\begin{equation}
  \label{eq:wall}
  \imag\left(\frac{t^2/2}{2\pi\ri t}\right) = \real t =
  0\;\Leftrightarrow \; \imag \tau = 0.
\end{equation}
In our discussion we have always assumed
\begin{equation}
  \label{eq:Hpos}
  \imag \tau > 0 \;\Leftrightarrow \; \real t >0,
\end{equation}
and the wall of marginal stability is the natural boundary of this
region.

\begin{figure}
  \centering
  \includegraphics[height=4cm]{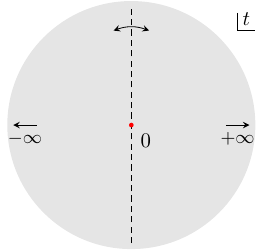}
  \caption{The moduli space and the possible wall of marginal
    stability of type IIA string compactified on $X_E$.}
  \label{fig:wall}
\end{figure}

It is possible to cross the wall and consider the region
\begin{equation}
  \label{eq:Hneg}
  \imag \tau < 0 \;\Leftrightarrow \; \real t < 0,
\end{equation}
as shown in Fig.~\ref{fig:wall}.  In the holomorphic limit, the free
energy $\CF_g(\tau)$ with $g\geq 2$ are polynomials of $E_2,E_4,E_6$.
We can define $E_{2k}(\tau)$ with $\tau$ in the lower half plane
$\CH_-$ and they are related to the usual $E_{2k}(\tau)$ in the upper
half plane $\CH_+$ by
\begin{equation}
  \label{eq:E2kneg}
  E_{2k}(\tau) = E_{2k}(-\tau).
\end{equation}
For instance
\begin{equation}
  E_2(\tau) = 
  1-24\sum_{n=1}^\infty\frac{q^n}{(1-q^n)^2} =
  1-24\sum_{n=1}^\infty\frac{1/q^{n}}{(1-1/q^{n})^2} = E_2(-\tau).
\end{equation}
Similarly, we can show
\begin{equation}
  E_4(\tau) = E_4(-\tau),\quad E_6(\tau) = E_6(-\tau).
\end{equation}
In addition, using
\begin{equation}
  1728\eta(\tau)^{24} = E_4(\tau)^3-E_6(\tau)^2,\quad \frac{\rd}{2\pi\ri\rd
    \tau}\log\eta(\tau) = \frac{E_2(\tau)}{24},
\end{equation}
we can also define
\begin{equation}
  \label{eq:etaneg}
  \eta(\tau) = \eta(-\tau),\quad \tau \in \CH_-.
\end{equation}
This allows us to define the free energy $\CF_g(\tau)$ with $g\geq 1$
in the regime \eqref{eq:Hneg}.  Given the simple identities
\eqref{eq:E2kneg}, \eqref{eq:etaneg}, we conclude that the instanton
sectors in the regime \eqref{eq:Hneg} are the same as in the regime
\eqref{eq:Hpos}, and the wall-crossing for the wall of marginal
stability \eqref{eq:wall} is trivial.

\section{Conclusion and discussion}
\label{sc:con}

In this paper we use the resurgence theory and the previous results on
the resurgent structure of topological string theory to study
non-perturbative corrections to the topological string model dual to
2d $U(N)$ Yang-Mills on torus, or equivalently the Gromov-Witten
theory of an elliptic curve.  We find closed form formulas for
instanton amplitudes up to arbitrary high instanton orders, based on
which we can write down the non-perturbative partition function with
contributions from all the real instantons which are in the resurgent
structure of the perturbative series.  This non-perturbative partition
function overcomes limitations of previous proposals and is in
particular real for positive modulus and string coupling.  We also
explore complex instantons and find two infinite towers as well as two
additional complex instantons.

Following \cite{Gu:2022sqc,Gu:2023mgf}, we expect that all the
instanton sectors, including the complex instantons, correspond to BPS
D-brane bound states in type II string on $X_E$.  We have also studied
the wall-crossing behavior of these states.  It would be interesting
to check and verify the spectrum of BPS states in the type II string.

One of the motivations to study the non-perturbative partition
function of topological string is to enable precise formulation of the
duality between 2d Yang-Mills on torus and topological string for
finite $N$, and more generally that of the OSV conjecture.  We wish to
make progress in this direction in the near future.

In this paper, we only restrict to the study of 2d Yang-Mills with
gauge group on torus.  For more generic target space of a Riemann
surface of arbitrary genus, there have recently been some progress on
the nature of the dual string description.  It would be interesting to
study the non-perturbative aspects of them.
Finally, it would also be interesting to study the dual string theory
for the 2d Yang-Mills of other gauge groups, for instance of the $B$,
$C$, $D$ types in the large $N$ limit. 

\section*{Acknowledgement}

We would like to thank Marcos Marino for participation in the initial
stages of the project, and for careful reading of the manuscript.
J.G. is supported by the startup funding No.~4007022316 of the
Southeast University, and the National Natural Science Foundation of 
China (General Program) funding No.~12375062.  X.W. is supported by
the National Natural Science Foundation of China Grants No.12247103.


\bibliographystyle{JHEP}
\bibliography{2dYM}  

\end{document}